\documentclass[12pt]{article}
\usepackage[left=2cm,right=3cm,top=2cm,bottom=2cm,bindingoffset=0cm]{geometry}
\usepackage{bbold}
\usepackage{amsmath,amsfonts,amssymb,amsthm,bm,slashed,bbm,mathrsfs}
\usepackage{graphicx,color}
\renewcommand{\Ref}[1]{(\ref{#1})}
\newcommand{\eq}[2]{\begin{align}\label{#1}#2\end{align}}

\newcommand{\beao}{\begin{eqnarray*}}
\newcommand{\eeao}{\end{eqnarray*}}
\newcommand{\be}{\begin{equation}}
\newcommand{\ee}{\end{equation}}
\newcommand{\bea}{\begin{eqnarray}}
\newcommand{\eea}{\end{eqnarray}}
\newcommand{\beq}{\begin{eqnarray}}
\newcommand{\eeq}{\end{eqnarray}}
\newcommand{\nn}{\nonumber}

\newcommand{\pa}{\partial}

\newcommand{\Ga}{\Gamma}

\newif\ifshowpictures
\showpicturestrue

\begin{document}

\title{$A_0$--condensation  in quark-gluon  plasma  with finite baryon  density
}
\author{M. Bordag\thanks{bordag@uni-leipzig.de}\\
{\small Institute for Theoretical Physics, Universit{\"a}t Leipzig}\\
 V. Skalozub\thanks{e-mail: Skalozubv@daad-alumni.de}\\
{\small Oles Honchar Dnipro National University, 49010 Dnipro, Ukraine}}

\date{Sept. 18, 2021}
\maketitle\thispagestyle{empty}

%
\begin{abstract}
In the present paper, we return to  the problem of  spontaneous generation of the $A_0$-background field in QCD at finite temperature and  a quark chemical potential, $\mu$.  On the lattice, this problem was studied by different approaches  where an analytic continuation to the imaginary potential $i \mu$ has been used. Here we consider  both,   real and imaginary chemical potential,
 analytically within the two-loop gauge-fixing  independent effective potential $W_{eff.}$. We realize the gauge independence in to ways: 1) on the base of Nielsen's identity and 2) expressing the potential in terms of Polyakov's loop. Firstly   we reproduce the known expressions in terms of Bernoulli's polynomials for the gluons and quarks. Then, we calculate the $\mu$-dependence, either for small $\mu$ as expansion or  numerically for finite $\mu$,  real and  imaginary.

One result is that the chemical potential only weakly changes the values of the condensate fields, but quite strongly deepens the minima of the effective potential.
We investigate the dependence of Polyakov's loop in the minimum of the effective potential, thermodynamic pressure and  Debye's  mass on the chemical potential. Comparisons with other results are given.
\end{abstract}
\section{\label{T1}Introduction}
Investigations of the deconfinement phase transition (DPT) and
quark-gluon plasma (QGP)   are in the center of modern high energy
physics.
One way to access these problems is the consideration of vacuum states in the presence of constant background fields.
For instance, a constant $A_0$-background allows  for an investigation of the deconfinement phase transition 
in terms of the average of the Polyakov loop,
\eq{PL}{ 
L=   \frac13 {\rm Tr}\  P \exp ( i g \int_0^\beta d x_4 ~A_0(x_4, \vec{x})),
}
(in $SU(3)$) as an order parameter. At finite temperature, $A_0$ cannot be zero and for reasons of gauge invariance there is a number of equivalent choices which provide equivalent vacuua. These states belong to the center $Z_3$ of the gauge group and the properties of, say the effective potential,
\eq{Z}{ Z(\mu) &= \int DA_\mu D\bar\psi D\psi \ \exp
	\left[\int_0^\beta d\tau \int dx\left(
	-\frac14 (F^a_{\mu\nu})^2+\bar\psi(i\ \slashed D +i\gamma^4 \mu)\psi   \right)\right],
}
or the Polyakov loop, \Ref{PL}, under the corresponding transformations, allow to discriminate confined and deconfined states. These ideas were discussed in literature many times beginning with  \cite{robe86-275-734} and evolved over the decades.

On the one loop level, the free energy $F=-\beta \ln Z $, belonging to \Ref{Z}, will be the blackbody radiation of the free gluon and quark fields. It has the known $Z_3$-structure for the gluon sector, which is broken in the quark sector. If including the next loop, it was found in the early 90-ies that there are non-trivial, lower lying  minima. These imply a spontaneous generation of the $A_0$-background, or in other words, a condensate. This is similar to the Savvidy vacuum (for a recent review see \cite{savv20-80-165}). While the magnetic case is plagued by instabilities, the $A_0$-background is stable.

It is common knowledge that the $A_0$-background appears at finite temperature. Due to the periodicity of the fields, it cannot be gauged away and the matrix $A_0=\sum_{a=1}^{N^2-1}A_0^a\,  t^a$  (for $SU(N)$) can be chosen to be diagonal (and traceless). The chemical potential $\mu$ appears as a shift, $A_0 \to A_0+i\frac{\mu}{g}\mathbb{1}$   proportional to the unit matrix, with imaginary chemical potential $\theta$, $i\mu\equiv \theta$. In the perturbative approach, on the one loop level this picture was understood in the early 80-ies, for instance in the papers \cite{weis81-24-475}, \cite{weis82-25-2667}, \cite{robe86-275-734} (see also the review \cite{gros81-53-43}). The two loop contribution was calculated in the early 90-ies, for gluons in \cite{bely90-45-355}, \cite{enqv90-47-291}, using the so-called 'mesonic (charged) basis'. In \cite{skal94-57-324}, overcoming some shortcomings in the literature, it was shown that in two loops the effective potential has non-trivial minima $A_0^3=\frac{g(3-\xi)}{4\pi\beta},~A_0^8=0$ in the basic sector.
The contribution of quarks in two loops was calculated therein. It was found that the minima known from the  gluonic sector  stay in place and that the effective potential gets lowered.

These results are gauge dependent which, naturally, caused questions. In \cite{skal94-50-1150}, the gauge dependence of the condensation was shown to satisfy the Nielsen identities. Recently, this topic was reconsidered in \cite{skal2006.05737}, showing that the result, expressed in terms of the eigenvalues of the Polyakov loop, is gauge invariant. Also, it was found that this is equivalent to take the gauge $\xi=-1$.
Another, independent, approach is the use of constraint potentials where the condensation shows up in three loop approximation. For a recent account see  \cite{kort20-101-094025} and literature cited therein.

In the case of quark chemical  potential $\mu$ the situation is not so well studied. Mainly, because the partition function becomes complex and  special considerations are needed. First of all this concerns the lattice calculations which must be  real basically and therefore an analytic continuation from imaginary values of $\mu$ to real values is applied.
These peculiarities are widely discussed in the literature for many years. In fact, they are both theoretical and
computational ones. Some new results and current references  can be found, for instance, in \cite{bori20-28-20}, \cite{bori21-965-115332}, where the results on condensation of $A_0^3 $ and $A_0^8$ in the   presence of $\mu$  are obtained in analytic and numeric lattice calculations based on a  model of  interacting Polyakov's loops. These are in agreement and correspondence with the ones obtained below.

In the present paper we investigate in detail how a chemical potential, real and imaginary, influences these features. We repeat some known results and present some new ones. For instance, using explicit analytic expressions  we discuss for the condensates the relation between imaginary and real chemical potential  which might be of relevance for the extrapolation of lattice results.

We also derive an expansion parameter which regulates an applicability of the  loop expansion theory for the effective potential $W_{eff}(A_0)$  at high temperature, $r = \frac{g^2}{8 \pi^2}$, which  is sufficiently small for values of coupling close to the confinement temperature. Using the constructed gauge-fixing independent effective potential $W_{L}(A_0^{cl})$, we calculate a thermodynamic pressure of plasma and the screening Debye's mass of gluons at the considered environment.
The  paper is organized  as follows. In the next section we provide the necessary information about the  Lagrangian, the notations  and the results  of the two-loop free energy in the  background potentials $A_0^3$ and $A_0^8$.   We also include the  quark chemical potential $\mu$.
In sect. \ref{T3}, for $\mu = 0$, we investigate gauge-fixing dependence of the effective potential within the Nielsen identity approach and the effective potential of order parameter and derive the gauge invariant effective potential.
In sects. 4 and 5  we consider in detail the dependence of the condensation on a chemical potential, both, real and imaginary.
In the last section we discuss  the results and give some outlook.
Appendix A is devoted to the derivation of $\mu$ dependent effective potential in terms of Bernoulli's polynomials.  Appendix B contains Feynman's rules and the definition of the periodic Bernoulli polynomials.

\section{\label{T2}Two-loop free energy}
In this section we remind the  expressions for the effective action in one and two loop order in a constant $A_0$--background, which are known so far, and discuss the inclusion of a finite chemical potential $\mu$ in the quark sector.
The starting point is
the QCD Lagrangian {with its gluon fields, $G^a_{\mu\nu}$, ghost fields, $\chi$ and quark fields, $\Psi ^a$,}
 in the background $R_\xi$  gauge,  Euclidean space-time and the presence of a chemical potential  $\mu$,
\eq{Lagrangian}{ L& = \frac{1}{4} (G^a_{\mu\nu})^2 + \frac{1}{2 \xi} ( D^B_\mu Q^a_\mu)^2 + \bar{\chi} D^B_\mu D_\mu \chi
 \\ \nn
&~~~~+  \bar{\Psi}^a (\gamma_\mu \partial_\mu + i m )\Psi ^a  + i g \bar{\Psi}^a \gamma_\mu ( A^c_\mu + Q^c_\mu ) ( t^c )^a_b \Psi^b - \mu \bar{\Psi}^a \gamma_4 \Psi ^a  ,
}
where $( t^c )^a_b$ are generators of $SU(3)$ group,
\eq{A0}{
	A^c_\mu = \delta_{\mu 0} ( \delta^{c 3} A_0^3 + \delta^{c 8} A_0^8 )
}
are the components of the background field and
\eq{fs}{ G^a_{\mu\nu} &= ( D^B_\mu)^{a b} Q^b_\nu - ( D^B_\nu)^{a b} Q^b_\mu  - g f^{a b c} Q^b_\mu Q^c_\nu,  \\\nn
 (D^B_\mu)^{a b}& = \delta^{a b} \partial_\mu + g f^{a b c} A^c_\mu,~~~ ( D_\mu)^{a b} = \delta^{a b} \partial_\mu + g f^{a b c} Q^c_\mu
}
holds.
It is useful to introduce a "charged (mesonic)  basis" for gluon fields \cite{bely90-45-355},\cite{skal94-50-1150},
\bea    \pi^{\pm}_\mu &=& \frac{1}{\sqrt{2}}( Q^1_\mu \pm i Q^2_\mu), ~~\pi^0_\mu = Q^3_\mu, \nn \\
K^{\pm}_\mu &=& \frac{1}{\sqrt{2}}( Q^4_\mu \pm i Q^5_\mu), \nn \\
 \bar{K}^{\pm}_\mu& =& \frac{1}{\sqrt{2}}( Q^6_\mu \pm i Q^7_\mu), ~\eta_\mu =  Q^8_\mu.     \label{fields} \eea
{These sets of  fields correspond to the $SU_I(2)$, $SU_U(2)$ and $SU_V(2)$ subgroups of the $SU(3)$ group. In terms of them the calculations and the representation of results are considerably simplified.}

In particular, in this basis the constant background potentials  $ A_0^3 $ and $ A_0^8 $  enter  the  Lagrangian, and the Feynman rules, in the form of a shift of the  fourth momentum component of  the color charged fields in \Ref{fields},
%
\eq{mod}{p_4  	\to 2\pi T\left( l + a_i\right),~~~(i=1,2,3),
}
with
\eq{mod1}{ a_i &= \frac{g}{2\pi T}   \left[
	\cos\left(\frac{\pi(i-1)}{3}\right)A_0^3
	+	\sin\left(\frac{\pi(i-1)}{3}\right)A_0^8   \right], &x=&\frac{g}{\pi T}A_0^3,
\\\nn &\equiv \frac12   \left[
\cos\left(\frac{\pi(i-1)}{3}\right)x
+	\sin\left(\frac{\pi(i-1)}{3}\right)y   \right], &y=&\frac{g}{\pi T}A_0^8.
}
The explicit expressions read
\eq{ai}{ a_1= \frac{x}{2},~~a_2= \frac{1}{4} ( x + \sqrt{3} y),~~a_3= \frac{1}{4} (- x + \sqrt{3} y).
}
Here $a_1$ is related to the $SU_I(2)$ color isotopic spin subgroup, and  $a_2$ and $a_3$ correspond to the $SU_U(2)$, $SU_V(2)$ ones.
The remaining three charged gluon lines follow with the property
\eq{ais}{a_{i+3}=-a_i.
}
The three quark lines undergo shifts
\eq{modq}{p_4  	\to 2\pi T\left( l + c_i\right)
}
with
\eq{modq1}{ c_i &= \frac{g}{2\pi T} \ \frac{1}{\sqrt{3}}  \left[
	\cos\left(\frac{\pi(2i-3/2)}{3}\right)A_0^3
	+	\sin\left(\frac{\pi(2i-3/2)}{3}\right)A_0^8   \right]+\frac12+i\tilde \mu,
\\\nn &\equiv \frac{1}{2\sqrt{3}}   \left[
	\cos\left(\frac{\pi(2i-3/2)}{3}\right)x
	+	\sin\left(\frac{\pi(2i-3/2)}{3}\right)y   \right]+\frac12+i\tilde \mu,
}
with the notation  $\tilde\mu=\mu/(2\pi T)$.
The explicit expressions are
\eq{ci}{ c_1= \frac14\left(x+\frac{y}{\sqrt{3}}+2\right)+i\tilde \mu,
	~~c_2= \frac14\left(-x+\frac{y}{\sqrt{3}}+2\right)+i\tilde \mu,
	~~c_3= -\frac{y}{2\sqrt{3}}+\frac12 +i\tilde \mu.
}
These parameters follow from the covariant derivative in the fundamental representation and are connected by $q_i=2c_i+1$ with those used in \cite{gros81-53-43}. The $c_i$ include the fermionic '+$\frac12$'. The relation to the parameters $a_i$, which follow from the adjoint representation, reads $a_1=c_1-c_2$, $a_2=c_1-c_3$, $a_3=c_2-c_3$. The parameters $a_i$ are physically more instructive than the differences $q_i-q_j$ used frequently in literature.
We mention the symmetry under rotation by $\pi/3$ in the $(A_0^3,A_0^8 )$ plane, or equivalently, in the $x,y$-plane, of the six frequency shifts  of the charged gluons. There is a corresponding symmetry for the quark lines, however with the double angle, $2\pi/3$. Also we mention that the sums of the parameters $a_i$ (accounting for \Ref{ais}) and of $c_i$ (except for the addendum $+\frac12+i\tilde\mu$) are zero.

The corresponding Feynman's rules  are the standard ones (see also, for example, equations on p. 356, right column, in \cite{bely90-45-355}, for their relations to the fields), with the only modification    by $i\mu$. We remind them in App. B.

The first  calculations of  the effective action with a constant addendum to $p_4$ were undertaken long ago in \cite{anis84-10-423} and \cite{weis82-25-2667}. The results for one and two loop levels,  formulated in terms of the Bernoulli polynomials, were  obtained in \cite{enqv90-47-291} and  \cite{bely90-45-355}, both for gluons. In \cite{skal94-57-324}, the two loop contribution from quarks was calculated. In \cite{skal94-50-1150} and \cite{skal94-9-4747} it was shown that the results are in agreement with  Nielsen's identities. More recently, these results were reviewed and reformulated in terms of the Polyakov loop  in \cite{skal2006.05737}.

The effective potential, including the  two loops contributions of gluons and quarks, can be written in the form
\eq{Weff}{W_{eff}&=W_g+N_fW_q,
}
as sum of the gluonic part (we follow \cite{skal94-9-4747}, where some misprints in earlier papers were corrected),
\eq{Wg}{   \beta^4 W_g&=
	\frac{4\pi^2}{3}\left[-\frac{1}{30}+\sum_{i=1}^3 B_4(a_i)\right]	
	+\frac{g^2}{2} \left\{\sum_{i=1}^3 \left[B_2^2(a_i)+2B_2(0)B_2(a_i)\right]\right.
\nn\\ &~~~~ +B_2(a_1)B_2(a_2)+B_2(a_2)B_2(a_3)+B_2(a_3)B_2(a_1))\Bigg\}
\nn\\
&~~~+\frac{g^2}{3}(1-\xi)
		\big\{ 	B_3(a_1) [ 2B_1(a_1)+B_1(a_2)-B_1(a_3)]+
\nn\\&		~~~~~~~~~~		B_3(a_2) [ 2B_1(a_2)+B_1(a_1)+B_1(a_3)]+
\nn\\&			~~~~~~~~~~		B_3(a_3) [ 2B_1(a_3)-B_1(a_1)+B_1(a_2)] \big\},
}
and the quark part (eq. (2) in \cite{skal94-57-324})
\bea \label{Wq} W_q  \frac{\beta^4}{N_f}& =& - \frac{4 \pi^2}{3} \sum_{i = 1}^3 B_4 ( c_i ) - \frac{1}{2} g^2 \Bigl( B_2(a_1) [ B_2(c_1) +  B_2(c_2)] \nn \\
&&+  B_2(a_1) [ B_2(c_1) +  B_2(c_3)] +  B_2(a_3) [ B_2(c_3) +  B_2(c_2)] \nn \\
&&-  \frac{1}{3} \sum_{i = 1}^3 [ B_2^2(c_i) - 2 B_2(i  \mu) B_2(c_i) ] - B_2(c_1) B_2(c_2) - B_2(c_2) B_2(c_3) - B_2(c_3) B_2(c_1)  \Bigr)  \nn \\
&&+ \frac{(\xi - 1)}{3} g^2 \Bigl[B_1(a_1)( B_3(c_1)-B_3(c_2)) \nn \\
&&+  B_1(a_2)( B_3(c_1)-B_3(c_3)) +  B_1(a_3) (B_3(c_2)-B_3(c_3)) \Bigr]. \eea
Here, $N_f$ is the number of quark flavors. We also take for simplicity $\mu$ to be the same for all flavors. 
It is seen that the one-loop contribution is additive and the two-loop part ($\sim g^2$) includes interference terms of the corresponding color subgroups.

Starting from here, the discussion in this section is for $\mu=0$. In the representations used in literature so far, the polynomials, initially defined in $0\le x\le 1$, were continued to $-1\le x\le0$ to obey $B_n(-x)=(-1)^nB_n(x)$ so that \Ref{ais} can be exploited. This way it was possible to account by three variables, $a_i$   ($i=1,2,3$), for all six charged gluon fields. In the present paper we use a modified scheme. We define, starting from the initial interval $0\le x\le 1$, the continuation of the Bernoulli polynomials by periodicity in the real part of the argument. The above formulas \Ref{Wg} stay in place which follows from the property $B_n(x)=(-1)^nB_n(1-x)$ of the Bernoulli polynomials together with the periodicity. In App. A we re-derive the Bernoulli polynomials, using a polylogarithm, where this property becomes obvious.

The effective potential  \Ref{Weff} depends explicitly on the gauge fixing parameter $\xi$.
This circumstance caused a large number of discussions about the physical meaning of $W_{eff}$ and the whole approach with the $A_0$-condensation. However, as it was shown in \cite{skal94-57-324} and \cite{skal94-9-4747}, this effective potential satisfies Nielsen's identity justifying its $\xi$-independence. This property is established along special characteristic orbits in the $(A_0,\xi)$ plane, where
the variation in gauge-fixing parameter is compensated by corresponding variation in fields.
At the same time, it is desirable to have a function which would be independent of $\xi$ and explicitly written in terms of gauge invariant
observable (Polyakov's loop, usually). This idea was expressed by Belyaev \cite{bely91-254-153} and we  call this object the "effective potential of order parameter", $W_L(A_0^{cl})$. However, Belyaev obtained the value  $A_0^{cl} = 0$ in the two-loop order calculations. Below, in sect. \ref{T3} we show that the choice $\xi=-1$ delivers effectively the gauge invariant results.

Consider the effective potential \Ref{Weff} as function  of $x$ and $y$. It has the known    minima. In the main topological sector,
\bea\label{main}  && 0 \leq a_1 \leq 1, ~~ 0 \leq a_2 \leq 1, ~~ - 1  \leq a_3 \leq 0, \nn \\
&&0 \leq c_1 \leq 1, ~~ 0 \leq c_2 \leq 1, ~~  0  \leq c_3 \leq 1,
\eea
there is one minimum, located at
\eq{min}{x^{(0)}_{{{min}}}& = \frac{3 - \xi}{4 \pi^2}\, g^2, ~~
	y^{(0)}_{min} = 0,
}
and the effective potential in this minimum reads
\eq{Wmin}  {  \beta^4 (W_q+N_fW_{gl})_{|{\rm min}}
& =
-\left(\frac{8}{45}+\frac{7}{60}N_f\right)\pi^2
+\left(\frac{1}{6}+\frac{5}{72}N_f\right)g^2
\\\nn&~~~~~-\left(1+\frac{1}{6}N_f\right)\frac{(3-\xi)^2}{32}\frac{g^4}{\pi^2}
+\dots,
}
where the first two terms correspond to the zero field case (these are present also without condensation and represent the blackbody radiation) and the last term is due to the condensation. It is negative and lowers the free energy.

The other minima are located on a circle with radius $x^{(0)}_{min}$ in the $(x,y)$-plane,
\eq{omin}{\left(\begin{array}{c} x_{min}^{(k)} \\y_{min}^{(k)} \end{array}\right)
&=
\left(\begin{array}{c} \cos(\frac{\pi}{3}k) \\\sin(\frac{\pi}{3}k) \end{array}\right)
x^{(0)}_{min}, ~~~(k=1,\dots,5),
}
and form a hexagon (in \cite{weis82-25-2667} this was mentioned as permutation symmetry). This is the way how, for the condensates, the above mentioned rotation symmetry is realized, whenever this is not obvious for the quarks since the $c_i$ rotate by $2\pi/3$. In Fig. \ref{fig:1}, in the right panel, we show these 6 minima, which appear in the two loop approximation.
{ For comparison, in the left panel we show the effective potential for $g=0$, where we have only the 'trivial' minimum at $x=0$.}
 By periodicity resulting from the Matsubara frequencies, all these minima are replicated on a super lattice given by the shifts ($n,m$ are integers)
\eq{slatt}{
\left(\begin{array}{c} \Delta x \\\Delta y \end{array}\right)   &=
\left(\begin{array}{c}2n+4m\\\frac{2}{\sqrt{3}}n \end{array}\right),
~~~~( \mbox{for gluons alone}),
\nn\\
\left(\begin{array}{c} \Delta x \\\Delta y \end{array}\right)   &=
\left(\begin{array}{c}2n+4m\\2 {\sqrt{3}} n \end{array}\right),
~~~~( \mbox{including quarks}).
}
The replications of the minima in the pure gluonic case realize the $Z_3$-symmetry.
All these minima have the same depth.  As we see, the depth of a minimum  increases with    increasing number of flavors.
As first mentioned in \cite{skal94-57-324}, the minima of the pure gluonic effective potential are not moved by the inclusion of the quarks. However, as can be seen from \Ref{slatt}, the replication happens in the y-direction with a three times larger period ($\sqrt{3}$ in place of $\frac{1}{\sqrt{3}}$) such that only a third of them remains.   This way, the $Z_3$-symmetry is broken by the quarks.  This difference is, of course, due to the different group representations; associated for the gluons and fundamental for the quarks. The above discussion illustrates the commonly expected (e.g., \cite{robe86-275-734}) breakdown of the $Z_3$-symmetry by the quarks   in loop expansion. We demonstrate this feature in Fig. \ref{fig:2}.

Finally in this section, we express the Polyakov loop \Ref{PL} in terms of the condensates. We have to insert, using the known SU(3) generators $t^c$,  \Ref{A0} and \Ref{mod1} into \Ref{PL} and to get $L$  in terms of $x$ and $y$,
\eq{L1}{ L &= \frac13\left(2\cos\left(\frac{\pi}{2}x\right) e^{i\frac{\pi}{2\sqrt{3}}y}+e^{-i\frac{\pi}{\sqrt{3}}y}\right).
}
For $x$ and $y$  we insert the condensates \Ref{omin} and \Ref{slatt} and arrive at
\eq{L2}{   L= \frac13\left(1+2\cos\left(\frac{\pi}{2}x^{(0)}_{min}\right)\right).
}
As will be shown in sect. \ref{T3}, one has to use \Ref{min} with $\xi=-1$.
It is seen that on the one-loop level, i.e., for $x^{(0)}_{min}=0$, we have $L=1$ and on the two loop level with $x^{(0)}_{min}$ given by \Ref{min}, a smaller value, $L\lesssim 1$. It is to be mentioned that expression \Ref{L2} remains unchanged under \Ref{omin} and \Ref{slatt}, i.e., it takes the same value in all minima{      ~related by these formulas}.

\ifshowpictures
\begin{figure}[h]
	\includegraphics[width=0.45\textwidth]{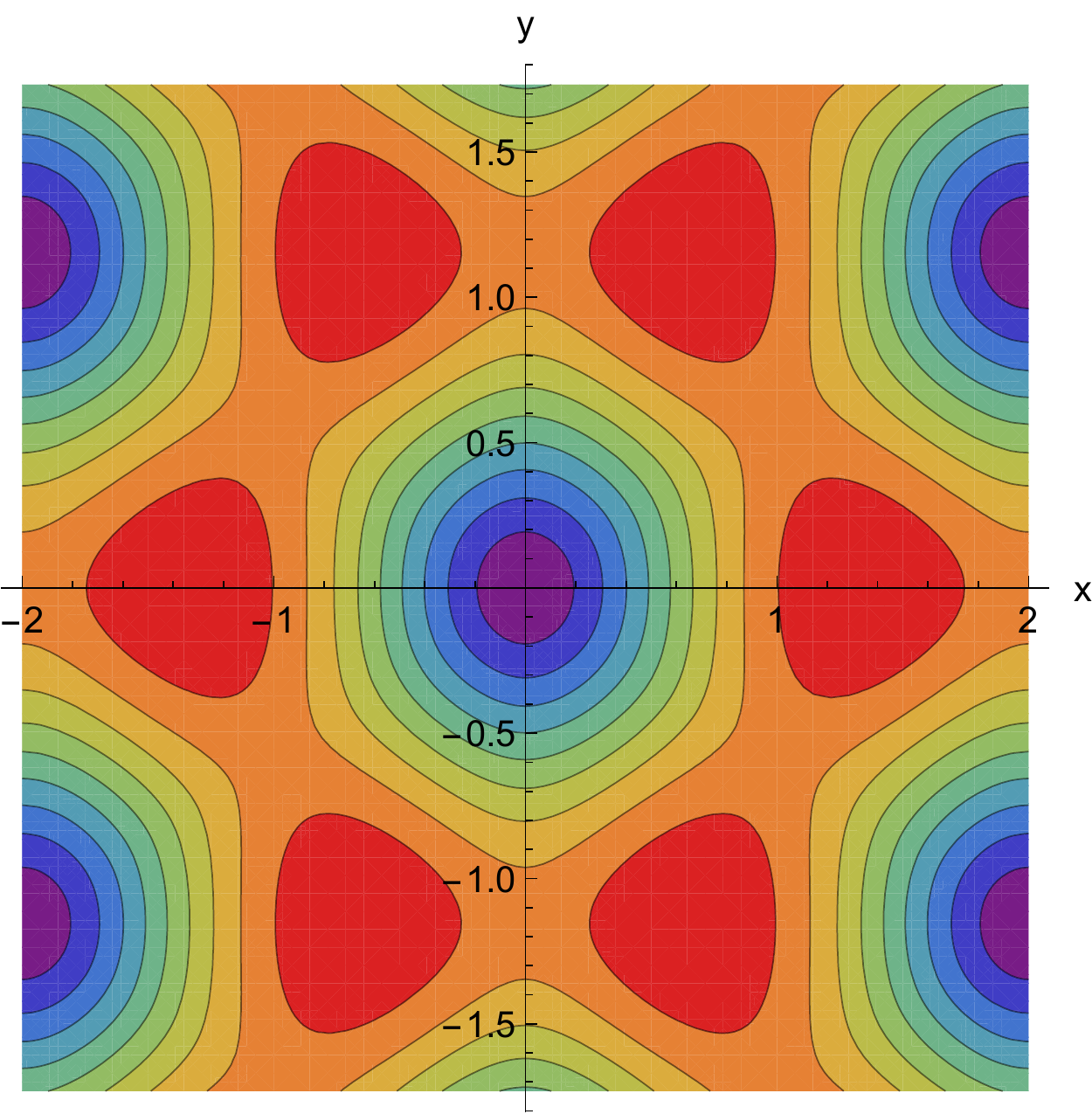}
		\includegraphics[width=0.53\textwidth]{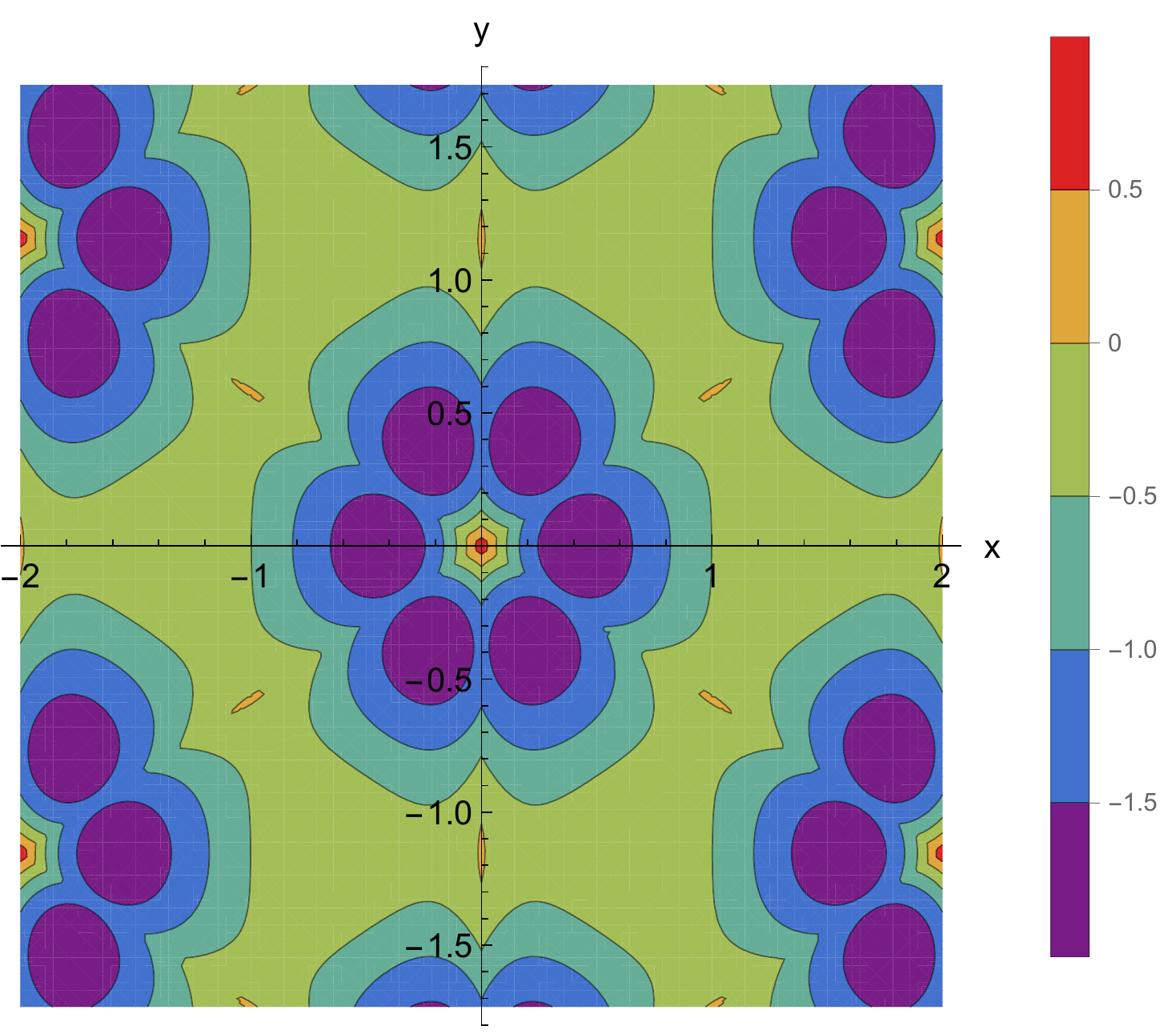}
		\caption{Contour  plot of the effective potential \Ref{Weff} for $N_f=3$, $\xi=-1$ and $g=0$ (left panel), $g=4$ (right panel).
			{The single minimum in the left panel is split into six in the right panel by the two loop contribution.}   }
\label{fig:1}\end{figure}

\begin{figure}[h]
	\includegraphics[width=0.45\textwidth]{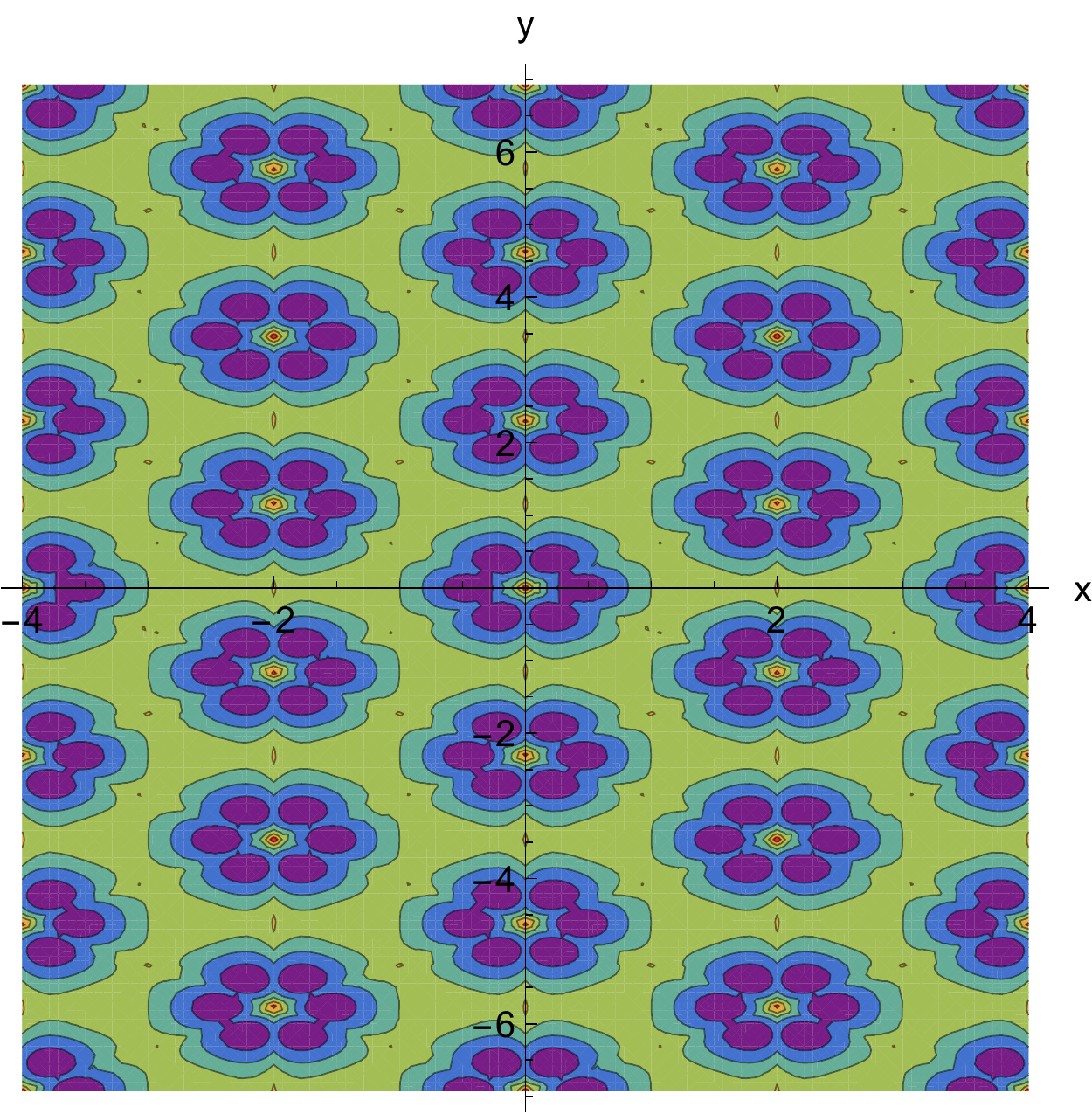}
	\includegraphics[width=0.52\textwidth]{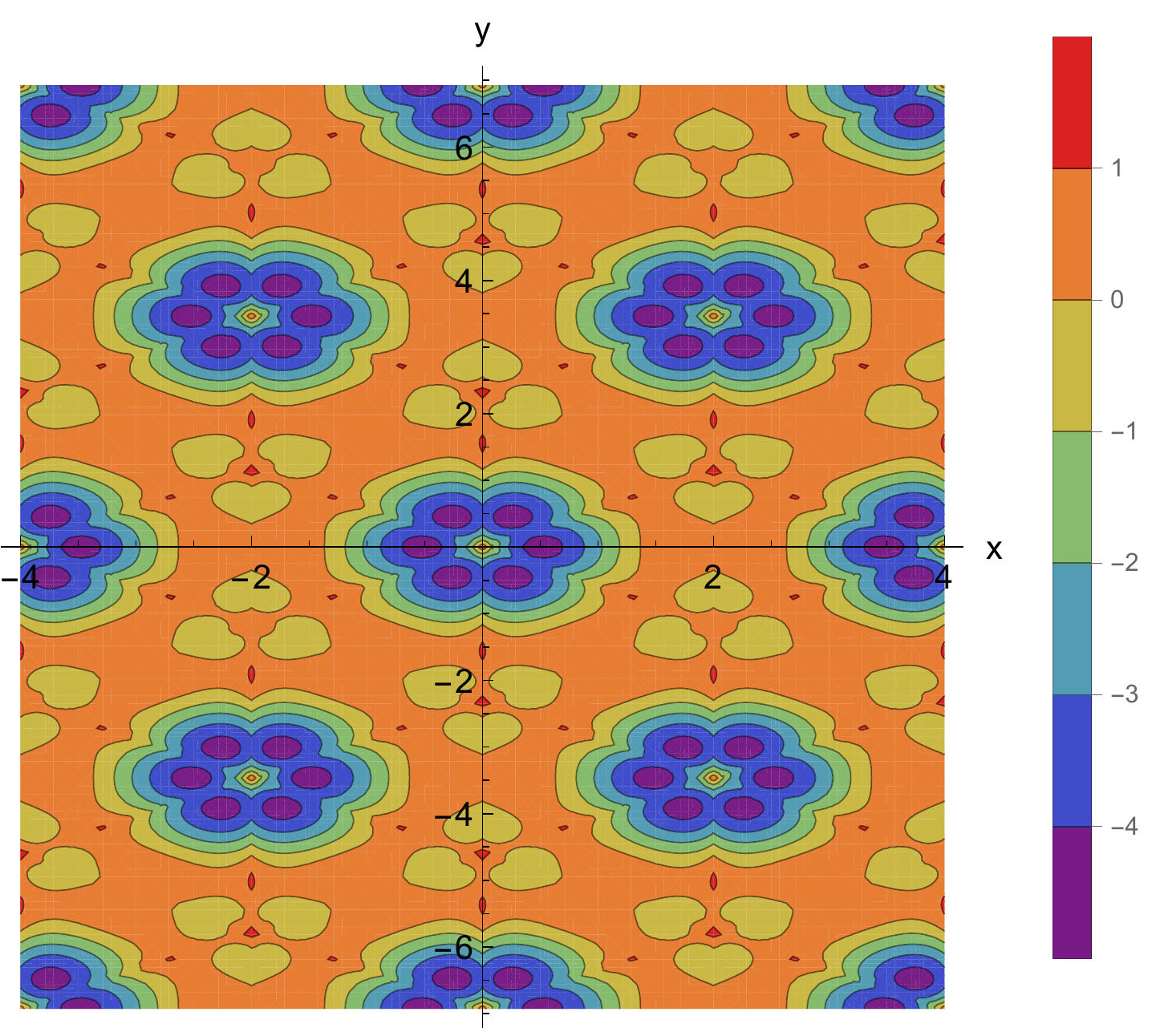}
	\caption{Contour plot of the effective potential \Ref{Weff} for $\xi=-1$, $g=4$ and  $N_f=0$ (left panel), $N_f=3$ (right panel). The replications according to  \Ref{slatt} are seen, especially the lifting of minima when including the quarks.}
\label{fig:2}\end{figure}

\fi

\section{\label{T3}On the gauge independence of the $A_0$-condensation}
When the minima of the effective potential in the $A_0$-background were found, their dependence on the gauge fixing parameter $\xi$, see  \Ref{min} and \Ref{Wmin}, raised questions whether these can be physical. There  were  two ways found to get an affirmative answer. One way involves   the idea to restrict the integration space  in the functional integral representing the effective action to such fields, which have a given value of the Polyakov loop \Ref{PL}. This approach rests on the ideas expressed in \cite{orai86-271-653} and is actively developed. We mention the very recent paper \cite{kort20-101-094025}. The other, much more direct way, was suggested in \cite{bely91-254-153}. In \cite{skal94-50-1150} it was related to the Nielsen identities, which are a special case of the Ward-Takahashi (Slavnov-Taylor) identities.
{The basic statement is that a change in the gauge fixing parameter can be compensated by a change in the background field, see, for instance, eq. (1.2) in \cite{niel75-101-173} or eq. (1.1) in \cite{kobe91-355-1}. The Nielsen identities cast this statement into formulas.}
We adopt this approach here, in a much simplified form, to get a self contained representation and clarify some deficiencies in the literature.

We start from the equivalent statement in \cite{bely91-254-153} that the effective potential in the minimum  (at zero sources) is a gauge invariant object. Now, in our case it depends on the background field  $A_0$ and on the gauge fixing parameter $\xi$. Since $A_0$ is a gauge potential, it is not physical (measurable) and may itself depend on $\xi$, $A_0=A_0(\xi)$. But, its dependence must be such that the effective potential is independent on $\xi$,
\eq{3.1}{\frac{d}{d\xi}\,W_{eff}(A_0(\xi),\xi)=0.
}
This way, we expect  a set of curves in the $(A_0,\xi)$-plane.
We mention that this statement was derived in \cite{bely91-254-153} using the Nielsen identities.

Now, in order to get the mentioned curves,  we apply \Ref{3.1} to the effective potential \Ref{Weff}. We restrict ourselves to the order $g^2$ in the perturbation expansion. Further, for simplicity, we restrict ourselves (for the rest of this section) to the $SU_I(2)$-case, where the effective potential reads
\eq{3.2}{\beta^4 W_{eff}&\equiv W^{(1)}+W^{(2)}
	\\\nn &=\frac{4\pi^2}{3}\left(-\frac{1}{90}+B_4(a)\right)
		+\frac{g^2}{2}\left[B_2(a)^2+2B_2(0)B_2(a)\right]
\\\nn  &	~~~~~~	+\frac{g^2}{3}(1-\xi)2B_1(a)B_3(a)
\\\nn &	-\frac{4\pi^2}{3}B_4(c) -\frac{g^2}{2}\left[2B_2(a)B_2(c)-\frac13\left(B_2(c)^2-2B_2(0)B_2(c)\right)\right]
\\\nn &	~~~~~~	-\frac{g^2}{3}(1-\xi)B_1(a)B_3(c).
}
For notation simplicity we consider $x=x(\xi)$ (in place of $A_0$), assume $\frac{\pa x} {\pa\xi}\sim g^2$ and remind, after \Ref{ai} and \Ref{ci}, the relations $a=\frac{x}{2}=\frac{gA_0^3}{2\pi T}$ and $c=\frac{x}{4}+\frac12$. Then we have, to order $g^2$, from \Ref{3.1},
\eq{3.3}{\frac{d}{d\xi}\,W_{eff}(x(\xi),\xi)=
	\frac{\pa W^{(2)}}{\pa \xi}+\frac{\pa W^{(1)}}{x}\frac{\pa x}{\pa\xi}=0.
}
Carrying out the derivatives, using \Ref{3.2}, we note
\eq{3.4}{ \frac{\pa W^{(2)}}{\pa \xi} &= -\frac{2g^2}{3}B_1(a)B_3(a)+\frac{g^2}{3}B_1(a)B_3(c),
\\ \nn      \frac{\pa W^{(1)}}{\pa x} &= \frac12 \frac{\pa W^{(1)}}{\pa a}+\frac14 \frac{\pa W^{(1)}}{\pa c},
}
and
\eq{3.5}{ \frac{\pa W^{(1)}}{\pa x} &= \frac{2\pi^2}{3}B_4'(a)-\frac{\pi^2}{3}B_4'(c).
}
Using the recursion relations, $B_n'(a)=n B_{n-1}(a)$, for the Bernoulli polynomials, and inserting into \Ref{3.3}, we arrive at
\eq{3.6}{ \left(\frac{\pa x}{\pa\xi}-\frac{g^2}{4\pi^2}B_1(a)\right)
	\left(B_3(a)-\frac12B_3(c)\right)=0.
}
The expression factorized and the equation for the curve in the $(A_0,\xi)$-plane is simply
\eq{3.7}{\frac{\pa x}{\pa\xi}=\frac{g^2}{4\pi^2}B_1(a).
}
To get the solution of this equation to order $g^2$, we may substitute in the right side for $a=\frac{x}{2}$ some tree value, $x_{cl}$, and integrate. As a result we get
\eq{3.8}{ x=x_{cl}+\frac{g^2}{4\pi^2}B_1\left(\frac{x_{cl}}{2}\right)(\xi-\zeta),
}
where $\zeta$ is an integration constant{, characterizing the gauge orbits. The solutions are parallel straight lines (at the given approximation). On the tree level we have $x=x_{cl}$ and, of course, no $\xi$-dependence.
More precisely, inserting this relation into \Ref{3.2} and expanding in a series of $g^2$, we could obtain a  $\xi$-independent expression for the effective potential as a function of the characterization $\zeta$ of the gauge  orbits.}

A second line of reasoning is based on the circumstance that the average $<L>$ of the Polyakov loop \Ref{PL} is gauge invariant. Following \cite{bely91-254-153}, we calculate this average perturbatively in order $g^2$ (one loop).
On the tree level we have $x_{cl}=\frac{g A_0^{cl}}{\pi  T}$ and
\eq{3.9}{<L>^{tree}=\cos\left(\frac{\pi}{2}x_{cl}\right)
}
holds. Including loop corrections, this relation turns into
\eq{3.10}{ <L>=\cos\left(\frac{\pi}{2}x\right)=<L>^{tree}+\Delta L.
}
For the correction $\Delta L$, from the known Feynman rules, the expression
\eq{3.11}{ \Delta L &=-\frac{g^2}{2 T}\sin\left(\frac{\pi}{2}x\right) I
}
with
\eq{3.12}{ I\equiv         T\sum_\ell \int\frac{d^3k}{(2\pi)^3}
	\left[
	\frac{1}{k_\ell(k_\ell^2+\vec k^2)}-(1-\xi)	\frac{k_\ell}{(k_\ell^2+\vec k^2)^2}
	\right]
}
and $k_\ell=2\pi T(\ell+a)$ follows. For details we refer to \cite{bely91-254-153}, eq. (9).
Such kinds of expression were calculated in the literature repeatedly. However, we found it instructing to give a version in terms of zeta functional regularization. Thus we consider the expressions
\eq{3.13}{ I=I_1-(1-\xi)I_2
}
with
\eq{3.14}{ I_1 &=\int_0^\infty \frac{dt}{t}\,\frac{t^{s+1}}{\Ga(s+1)}
	 T\sum_\ell \int\frac{d^3k}{(2\pi)^3}
	 \frac{1}{k_\ell}\, e^{-t(k_\ell^2+\vec k^2)},
	\\	
	 I_2 &=\int_0^\infty \frac{dt}{t}\,\frac{t^{s+2}}{\Ga(s+2)}
	T\sum_\ell \int\frac{d^3k}{(2\pi)^3}
	 {k_\ell}\, e^{-t(k_\ell^2+\vec k^2)},
}
with $s=0$ at the end. We integrate over the spatial momenta and, in the first integral, we integrate in $t$  by parts. For the factor $k_\ell$ in front of the exponential we use a derivative with respect to $a$. We arrive at
%
\eq{3.16}{  I_1=-\frac{1}{4\pi T}\,\frac{\pa}{\pa a}
		\int_0^\infty \frac{dt}{t}\,\frac{t^{s-1/2}}{(4\pi)^{3/2}\Ga(s+1)(s-\frac12)}
	\,T\sum_\ell\, e^{-t k_\ell^2}
}
and
\eq{3.17}{I_2=\frac{\Ga(s+1)(s+\frac12)}{\Ga(s+2)}I_1=\frac12 I_1 ~~~({\rm at}~s=0).
}
This way, from \Ref{3.13} we get
\eq{3.15}{ I=\left(1-(1-\xi)\frac12\right)I_1 = \frac{1+\xi}{2}I_1.
}
Further we apply Poisson re-summation to $I_1$,
\eq{3.18}{ I_1=-\frac{1}{4\pi T}\,\frac{\pa}{\pa a}
	\int_0^\infty \frac{dt}{t}\,\frac{t^{s-1}}{(4\pi)^{2}\Ga(s+1)(s-\frac12)}
	\sum_N \cos(2\pi a N)\,e^{-N^2/(4T^2t)}.
}
Further we drop the $T=0$ contribution (this is $N=0$) and can now put $s=0$. We arrive at
\eq{3.19}{I_1=-\frac{1}{4\pi T}\frac{\pa}{\pa a}
	\frac{1}{4\pi^2}\int_0^\infty\frac{dt}{t^2}\sum_{N=1}^\infty\cos(2\pi a N)
		\,e^{-N^2/(4T^2t)}.
}
Carrying out the $t$-integration, we may use the known formula
\eq{3.20}{\sum_{N=1}^\infty\frac{\cos(2\pi a N)}{N^2}
		=\frac12\left(Li_2\left(e^{i 2\pi a} 	\right)
		+Li_2\left(e^{-i 2\pi a}\right)
		\right)
		=\pi^2 B_2(a),
}
where $B_2(a)$ is the periodic continuation of a Bernoulli polynomial. Inserting into \Ref{3.18} gives
\eq{3.21}{ I_1=\frac{T}{4\pi }\,\frac{\pa}{\pa a}B_2(a)
	=\frac{T}{2\pi}B_1(a).
}
Finally, inserting into \Ref{3.11} and using \Ref{3.15} and \Ref{3.17} we come to
\eq{3.22}{\Delta_T L=
	-\frac{g^2}{8\pi}(1+\xi)\sin\left(\frac{\pi}{2}x\right)B_1(a)
}
and with \Ref{3.10} to
\eq{3.23}{<L>&= \cos\left(\frac{\pi}{2}x_{cl}\right)
	-\frac{g^2}{8\pi}   \sin\left(\frac{\pi}{2}x\right)
	B_1(a)(1+\xi),
\\\nn	&=\cos\left( \frac{\pi}{2}\left(x_{cl}+\frac{g^2}{4\pi^2}(1+\xi)B_1(a)\right)\right)+O(g^2)
}
Comparing with \Ref{3.8} we come to the conclusion
\eq{3.24}{ x=x_{cl}+ \frac{g^2}{4\pi^2}(1+\xi)B_1(a).
}

This formula can be compared with \Ref{3.8} and we find for the integration constant $\zeta= - 1$. Just for this value an arbitrary characteristic coincides with the curve corresponding physical "observable" $x_{cl}$. As a result, to express \Ref{3.2} in     terms of Polyakov's loop    we have  to set 
$\xi = - 1$ and $x = x_{cl}$ in the initial effective potential.

Another important consequence of \Ref{3.24}
is the ratio of two- and one-loop contributions $r = \frac{W^{(2)}}{W^{(1)}}$ determining the expansion parameter of the theory, $r = \frac{g^2}{8 \pi^2}$. It is small for sufficiently large coupling values $g \le 4-5$, and the two-loop effective potential \Ref{3.2} is  applicable till the confinement temperature where  Polyakov's loop turns  to zero.

{The relation between the $x$ and the $x_{cl}$ \Ref{3.24} can be immediately extended to $U-$ and $V-$ subgroups (these are the fields $K^{\pm}_\nu, \bar{K}^{\pm}_\nu $ in  \Ref{fields}). We have just to replace $a_1 = \frac{x}{2}$ by $a_2$ and  $a_3$ \Ref{ai},  correspondingly.  This possibility is due to the obvious fact that the calculation procedure resulting in  \Ref{3.24}  is independent  of the actual value of $a_1 = \frac{x}{2}$. More detailed derivation of \Ref{3.8}, (and hence \Ref{3.24})  can be found in \cite{skal94-50-1150} for $SU(3)$ case and in \cite{skal94-9-4747} for full QCD.     }

Thus, to derive the gauge-fixing independent effective potential expressed in terms of the Polyakov loop  we have to set in \Ref{Wg} $a_1 = a_1^{cl}, a_2 = a_2^{cl}, a_3 = a_3^{cl}$ and $\xi = - 1$.  From the obtained effective potential, the observable values of the condensed fields are determined from their minimum positions.

\section{\label{T4}$A_0$ condensation in  the presence of $\mu$}
In this section we investigate how the picture described in the preceding sections changes under inclusion of a chemical potential $\mu$, i.e., of a finite density of quarks. We present the results obtained in two forms, one for the gauge-fixing dependent effective potential, in order to verify Nielsen's identity. The other form is the the gauge-fixing independent effective potential $W_L(A_0^{cl}, \mu)$.

On the level of the Feynman rules, the inclusion of $\mu$ is shown in \Ref{modq1}. The structure of the Feynman diagrams remains unchanged and formally we arrive at the expressions \Ref{Wq}, where $\mu$ enters through the variables $c_i$, \Ref{modq1}. These formulas can be viewed as a kind of analytic continuation in the variables $A_0^{3,8}$, or $(x,y)$,  in \Ref{modq1}. The question arises whether this is justified, especially having in mind that the $B_n$ are not really polynomials. The answer is yes, as we demonstrate in Appendix A. For instance, it can be seen that the periodic continuation is to be done in the real part of the argument. In doing so, obviously the symmetries \Ref{omin} and \Ref{slatt} stay in place.

Including the chemical potential this way, we first consider  the $\mu$ dependence analytically.  As in sect. \ref{T2}, we find the minimum of the effective potential in the fundamental sector in the $(x,y)$-plane, namely its position and depth, and expand the resulting expressions for small $g$.
It must be mentioned that the effective potential \Ref{Weff}  is a fourth order polynomial in $x$ and $y$. Its roots are solutions of third order equations and contain third order square roots. First we do an expansion of these expressions in powers of $g$.

For the position of the minimum, as it turns out, we have now that  \Ref{min} is kept for  $N_f< 9$,
i.e., there is no dependence on $\mu$ for known values of flavors. 
However, the effective potential \Ref{Wmin} depends on $\mu$. It is a fourth order polynomial in $\mu$,
\eq{mmu1}{ &\beta^4(W_{gl}+N_f W_{q})_{|{\rm min}} =
\\\nn &~~~	 W_0	-i\frac{ g^2}{12}N_f\tilde\mu
		+\left( - 2\pi^2+\frac{11 g^2}{12}-\frac{ g^4(3-\xi)^2}{16\pi^2}\right)N_f\tilde\mu^2
		+i  g^2N_f\tilde\mu^3
		+(4\pi^2-g^2)N_f\tilde\mu^4,
}	
where $W_0$ is the same as \Ref{min}	(recall $\tilde\mu=\mu/(2\pi T)$).

We see that   the chemical potential, for small $\mu$,  does not change   the condensate, but deepens the minimum. There are odd powers of $\mu$ (and with them an imaginary part for real chemical potential). A plot of the real part of this potential is shown in Fig. \ref{fig:3a}.

Again we note that the chemical potential enters \Ref{Weff} through the arguments of Bernoulli's polynomials in \Ref{modq1}. This does not change the structure of the effective potential that satisfies Nielsen's identity derived in \cite{skal94-9-4747},  \cite{skal94-57-324}, and gauge invariance is preserved. In order to obtain the gauge-fixing independent effective potential corresponding to Polyakov's loop, we have to substitute $A_0 \to  A_0^{cl}$, $ \xi \to \zeta  = - 1$, as it is described above. These substitutions  are equivalent to put $\xi=-1$ in  \Ref{mmu1}. The emerging expression can be used for calculating thermodynamic pressure in the plasma which is determined as: $ p = - Re (\beta^4(W_q+W_{gl})_{|{\rm min}})$. Hence it follows  that chemical potential increases the pressure.

In  \Ref{L2} we have seen that the Polyakov loop \Ref{PL} depends of the condensate, see  \Ref{L1}. Following the $\mu$ independence of the condensates,   $L$ is also independent on $\mu$.
\ifshowpictures
\begin{figure}
	\includegraphics[width=0.45\textwidth]{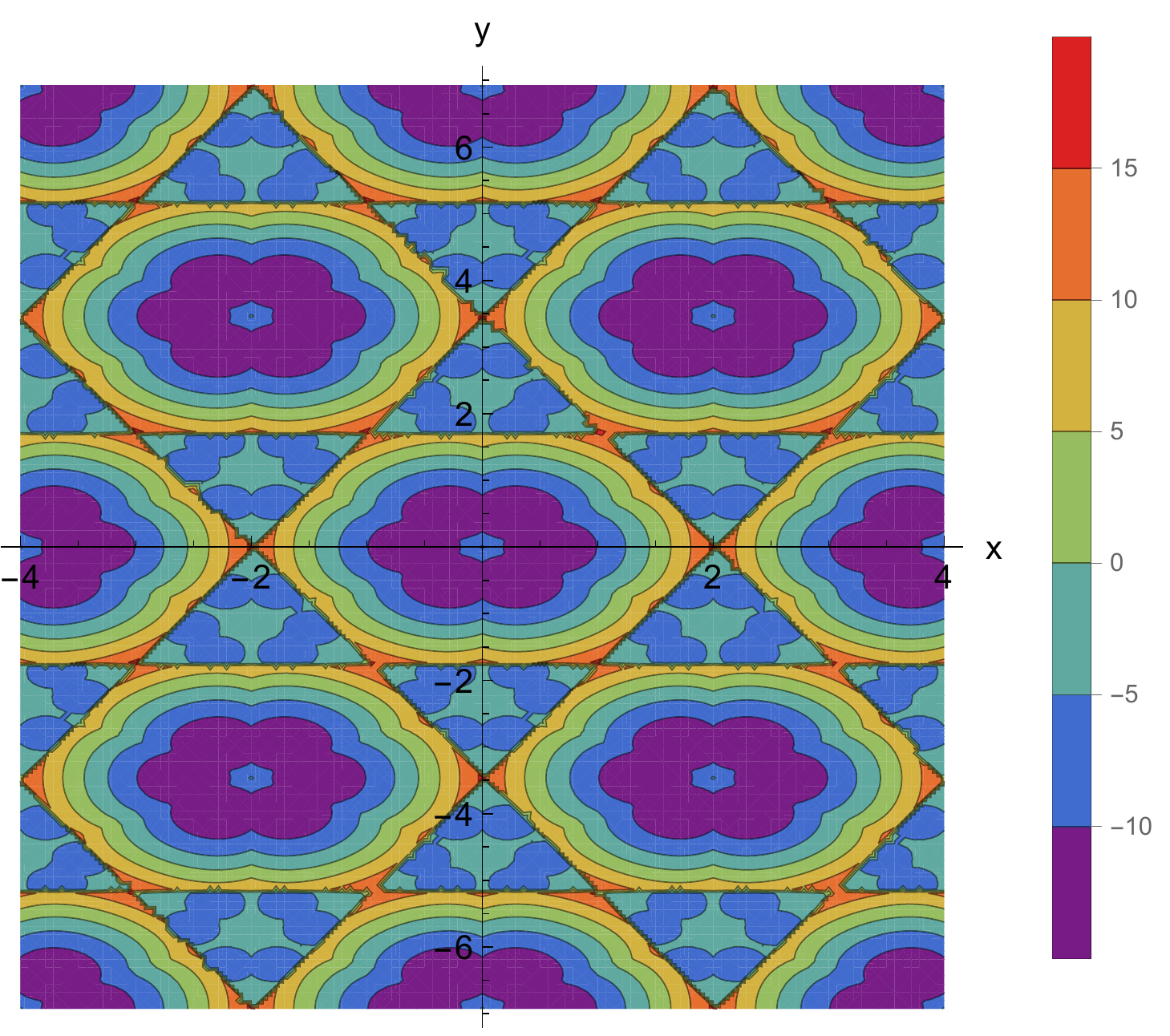}
	\includegraphics[width=0.45\textwidth]{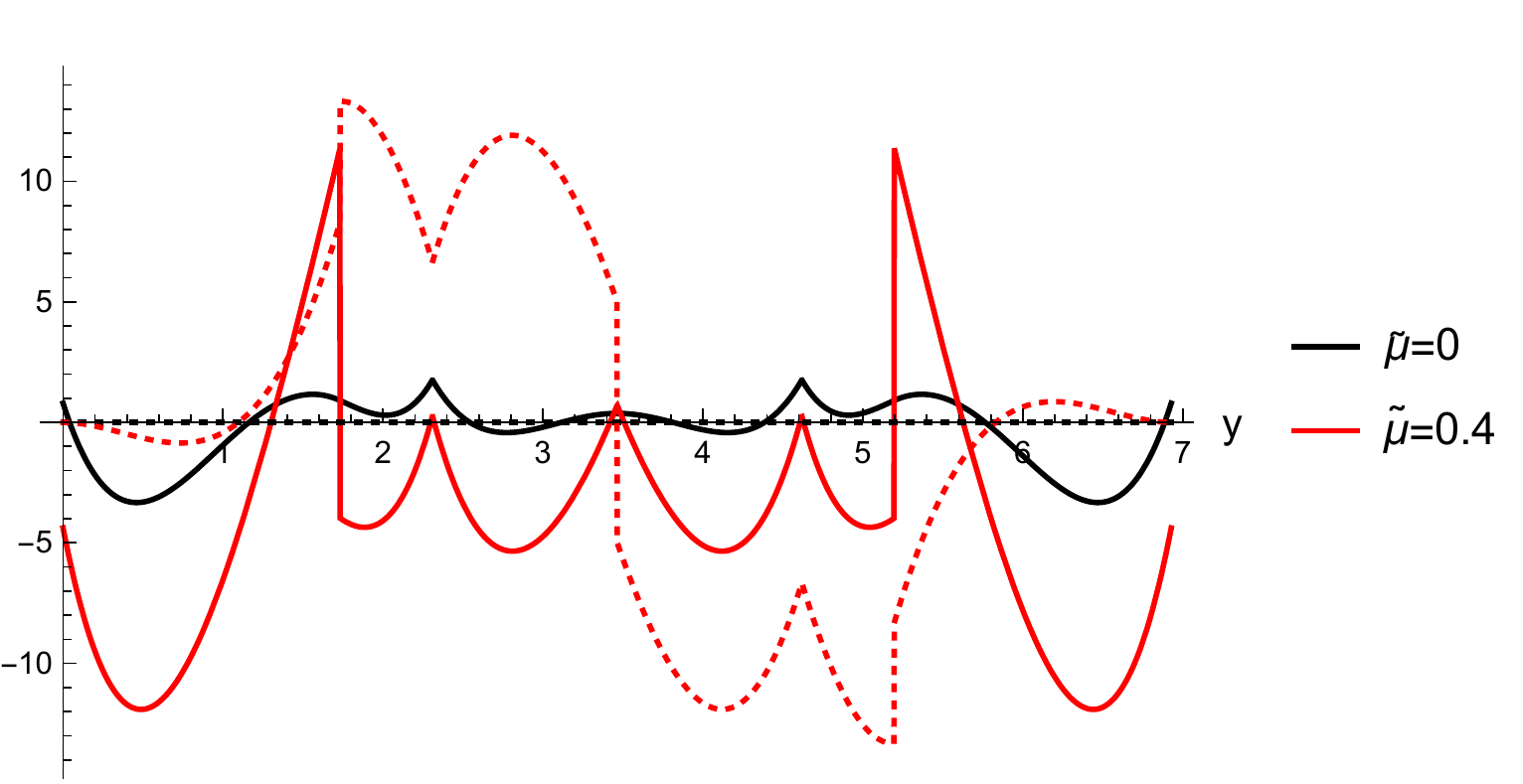}
	\caption{Contour plot (left panel) of the real part of the  effective potential with chemical potential $\tilde\mu=0.4$ and $g=4$, $N_f=3$, to be compared with Fig. \ref{fig:2} (right panel). Section at $x=0$ (right panel), to compare $\tilde\mu=0$ and $\tilde\mu>0$. The jumps result from the product of the odd-number polynomials in \Ref{Wq}. The dotted line is the imaginary part present for $\tilde\mu=0.4$.}
	\label{fig:3a}\end{figure}
\fi
\begin{figure}
		\includegraphics[width=0.45\textwidth]{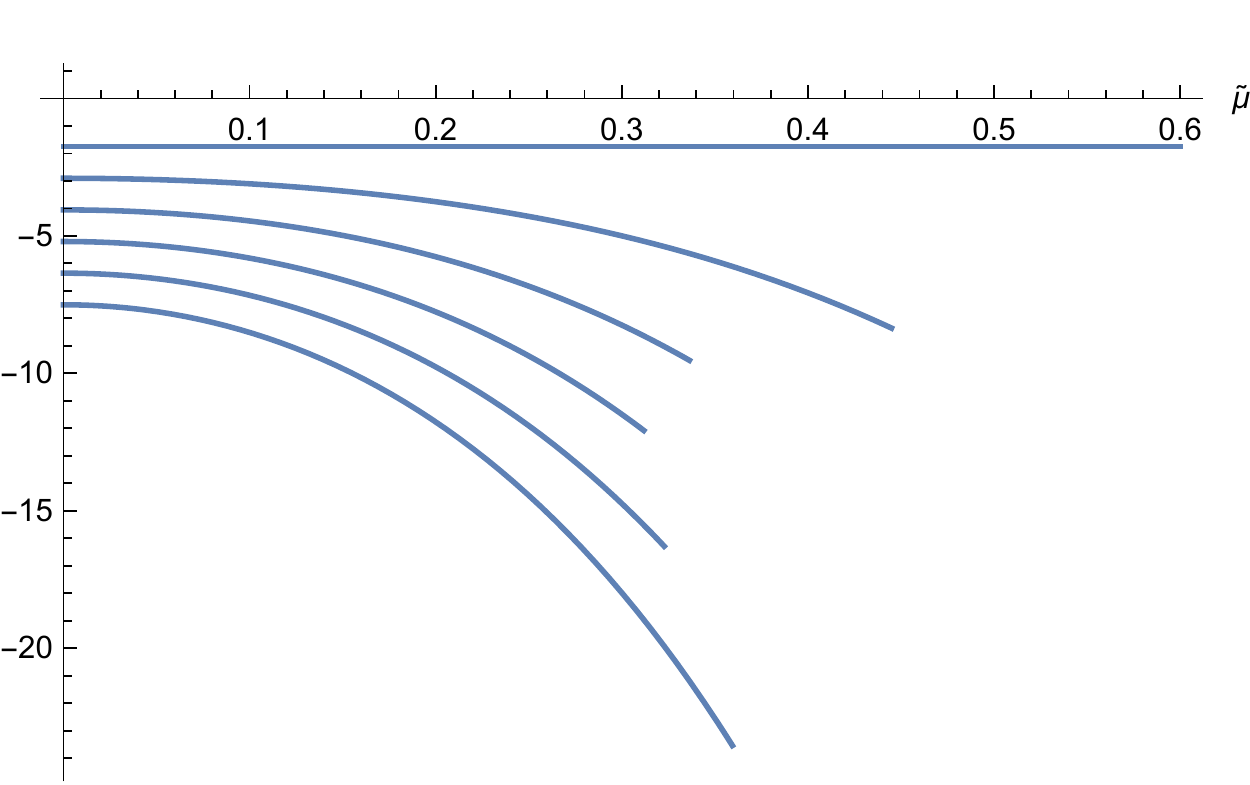} 
\caption{The effective potential \Ref{mmu1} in the minimum as function of $\tilde\mu$ for $N_f=0,1,...,5$ (from top to bottom). The curves terminate where the condensate ceases to be real.}
	\label{fig:Wmu}\end{figure}

{
We complete this section with the calculation of the Debye screening mass for neutral gluons.
 This mass is defined as
\eq{Dmass}{ m_D^2 &= \frac{d^2 W_{eff}}{d (A_0^3)^2 },
=\frac{g^2T^2}{\pi^2}\,\frac{d^2}{dx^2}\beta^4(W_{gl}+N_f W_q).
}
It is a function of the background, $x$ (we put $y=0$ for simplicity), $N_f$ and $\mu$.
 This function is a polynomial in $\mu$. For $x=0$ it reads
\eq{Md0}{m_{D,0}^2 &=\left(1+\left(\frac16+2\mu^2\right)\right)N_f g^2 T^2
	+\left(\frac{33}{16}-\xi+\frac{2-3\xi}{48}N_f+\frac{i \mu}{12}N_f+\frac{16-9\xi}{12}N_f\mu^2\right)\frac{g^4T^2}{\pi^2}.
}
In the minimum \Ref{min} of the effective potential the Debye mass is
\eq{Mdmin}{m_{D,min}^2 &=\left(1+\left(\frac16-2\mu^2\right)\right)N_f g^2 T^2
\\\nn&~~~	+\left(\frac{3(1-2\xi)}{16}+\frac{(2-3\xi)N_f}{48}+\frac{i\mu}{12}N_f+\frac{16-9\xi}{12}N_f\mu^2\right)
	\frac{g^4T^2}{\pi^2}.
}
The difference between them is $m_{D,0}^2-m_{D,min}^2=\frac{5(3-\xi)}{8}\frac{g^4T^2}{\pi^2}$.
As already, to obtain the Debye mass expressed in the terms of the Polyakov loop we have to set $\xi = - 1$ in the above expressions.
Finally, we mention the special case when  $\mu=0$, $N_f=0$,
\eq{md4}{m_{D,0}^2=g^2 T^2+\frac{49 g^4 T^2}{16 \pi ^2},~~~
	m_{D,min}^2=g^2 T^2+\frac{9 g^4 T^2}{16 \pi ^2},
}
which is the  pure gluonic case.
}

\section{\label{T4a}Condensation with imaginary chemical potential}
In this section we consider {\it imaginary} chemical potential. In the above formulas this  is reached by the substitution $\mu\to i\mu$ (and $\tilde\mu\to i \tilde\mu$). In that case the addendum to the frequency \Ref{modq} is real. As such, it fits to the periodic continuation of the Bernoulli polynomials and the effective potential will be periodic in this $\mu$ (with period 1 for $\tilde\mu$). In the perturbative results in Section \ref{T3} one has to do simply the mentioned substitution. The first order (in $\mu$) correction will dominate and its sign depends on the sign of $\mu$. The position of the minima shows a weak dependence on $\mu$. There is also a small contribution to $y$.  It is shown in  Fig. \ref{fig:3b}. It should be mentioned that this dependence is for finite $g$, in distinction from the preceding section where we considered the perturbative expansion in $g$.

We provide also some pictures of the effective potential for imaginary $\mu$, see Fig. \ref{fig:4}. As can be seen, the role of the minima changes as function of $\mu$, cycling with period of approximately $1/3$ (for $\tilde\mu$). We demonstrate this by plotting the depths of three consecutive minima   in  Fig. \ref{fig:5} (left panel).

\ifshowpictures
\begin{figure}
	\includegraphics[width=0.5\textwidth]{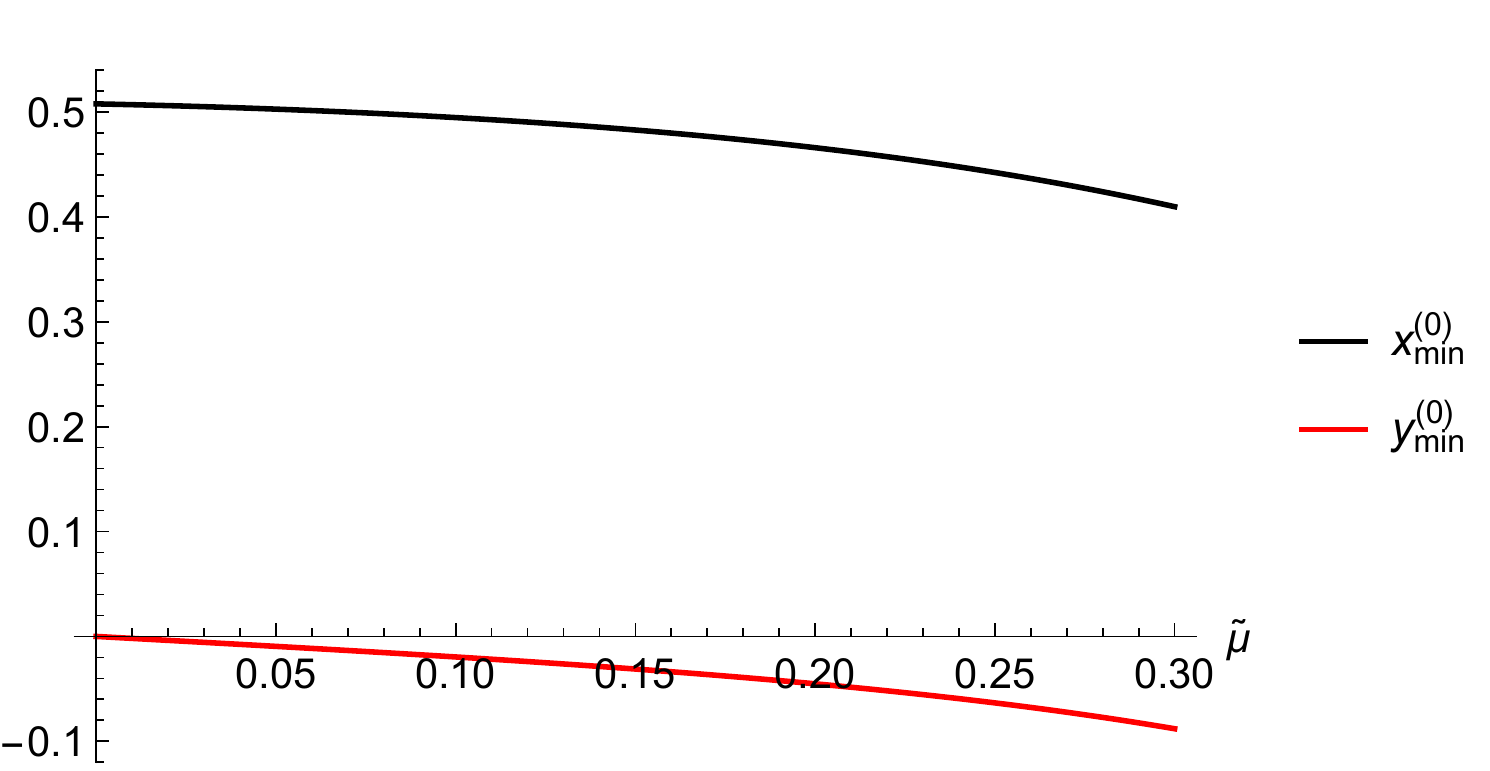}
	\caption{The (weak) dependence of the position of the minimum in the basic sector on imaginary chemical potential. Beyond the $\tilde\mu$ shown in the figure, this minimum disappears and another minimum becomes deeper (as demonstrated in Fig. \ref{fig:4}). The parameters are $g=4$, $N_f=3$.}
	\label{fig:3b}\end{figure}
\fi

The relation between real and imaginary potential is in the above formulas in complete analogy with the relation between trigonometric and hyperbolic functions. This can be most clearly seen by looking on the left side of  \Ref{A.8}, i.e., before the application of Jonqui\`{e}re inversion formula. Although this is on some stage nothing more than some analytic continuation, the results are quite different and hard to connect post factum.

\ifshowpictures\begin{figure}
	\includegraphics[width=0.29\textwidth]{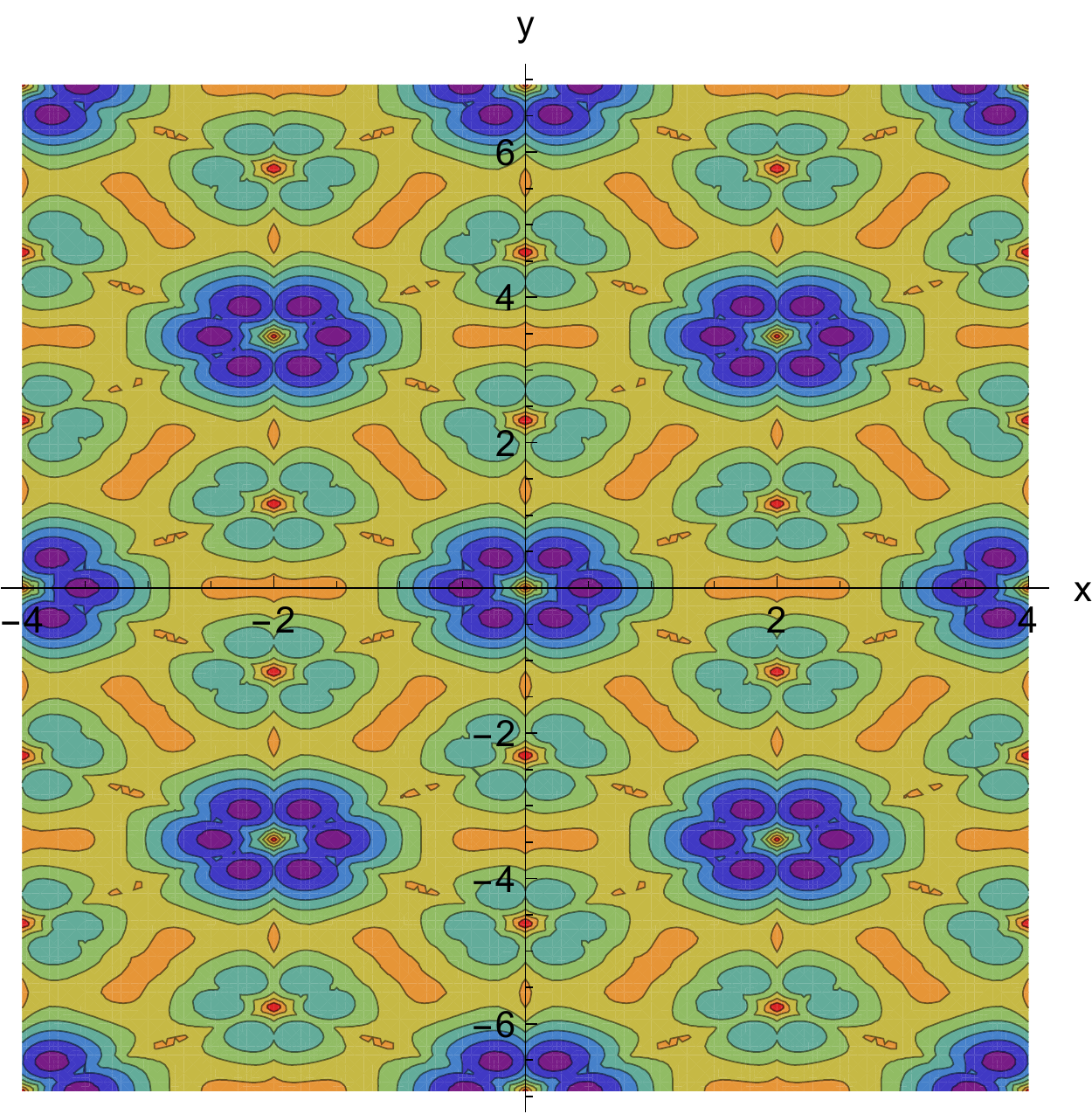}
	\includegraphics[width=0.29\textwidth]{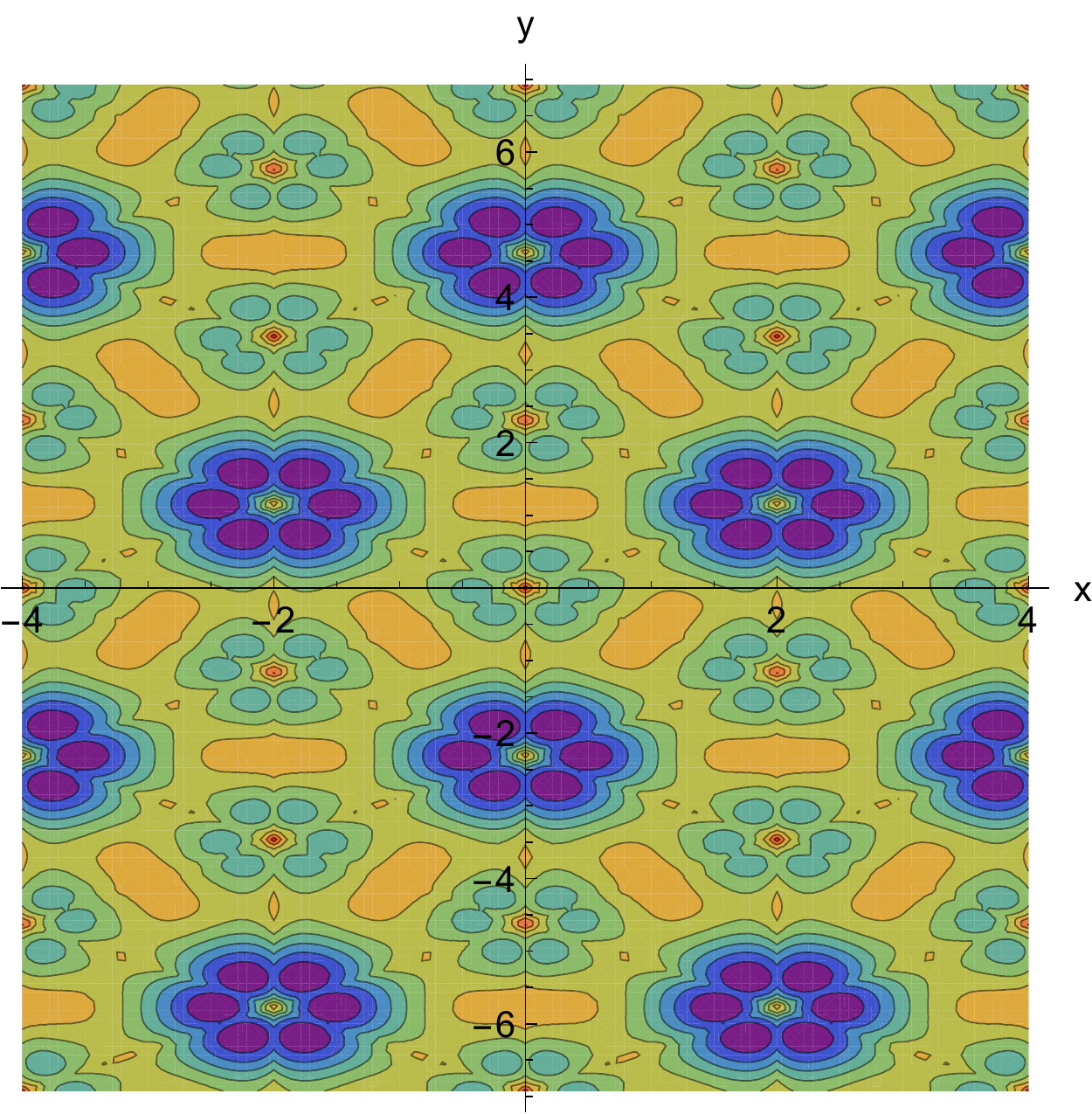}
	\includegraphics[width=0.34\textwidth]{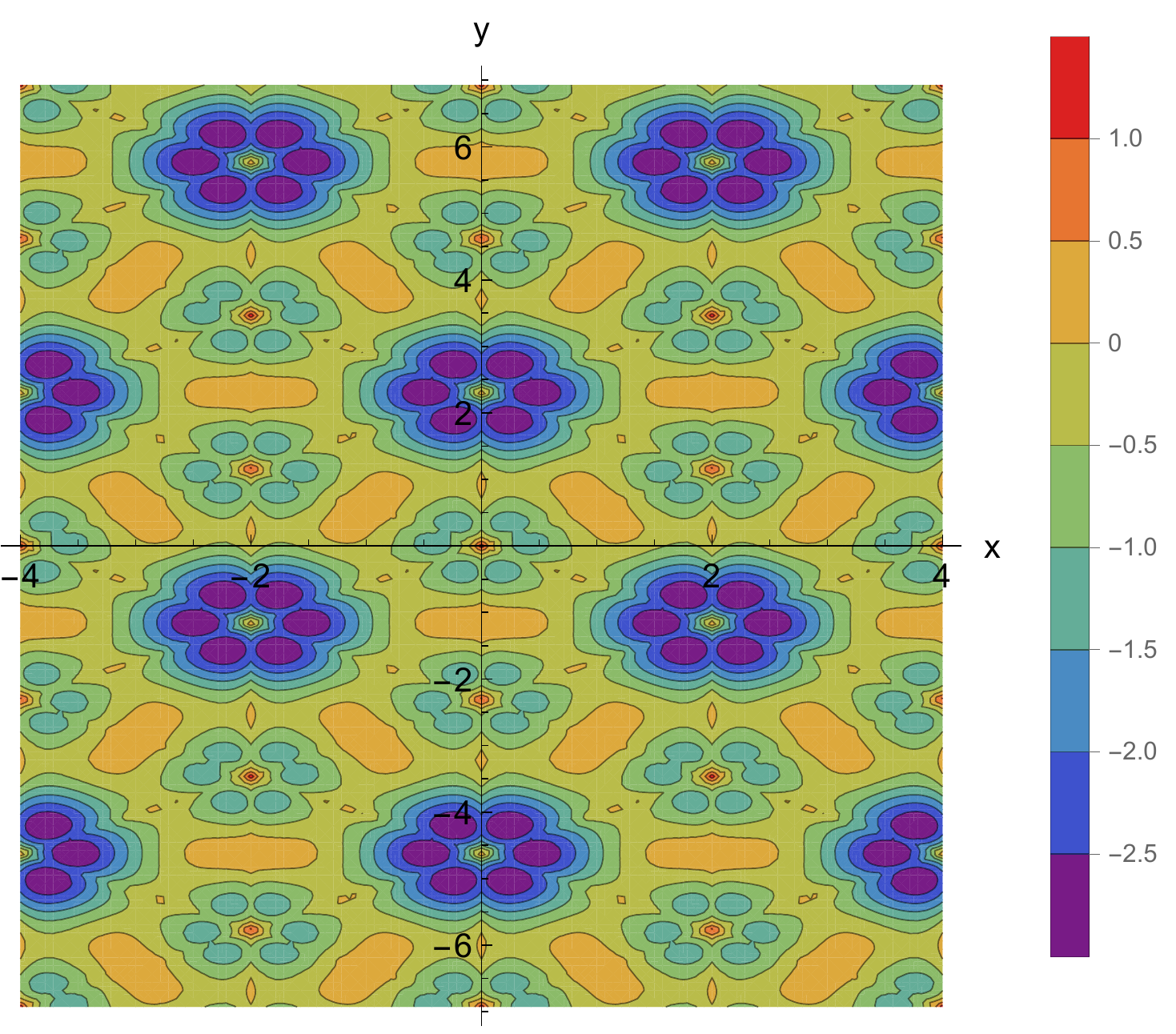}
	\caption{Contour plot   of the  effective potential with imaginary chemical potential $i\mu$ for  $g=4$, $N_f=3$. The dependence on $\mu$ is periodic and the panels are made with equal spaced $\mu$ within one period, $\tilde\mu=0$, $\tilde\mu=\frac13$, $\tilde\mu=\frac23$. The left panel is the same as the right panel in Fig. \ref{fig:2}.}
\label{fig:4}\end{figure}

\begin{figure}
	\includegraphics[width=0.45\textwidth]{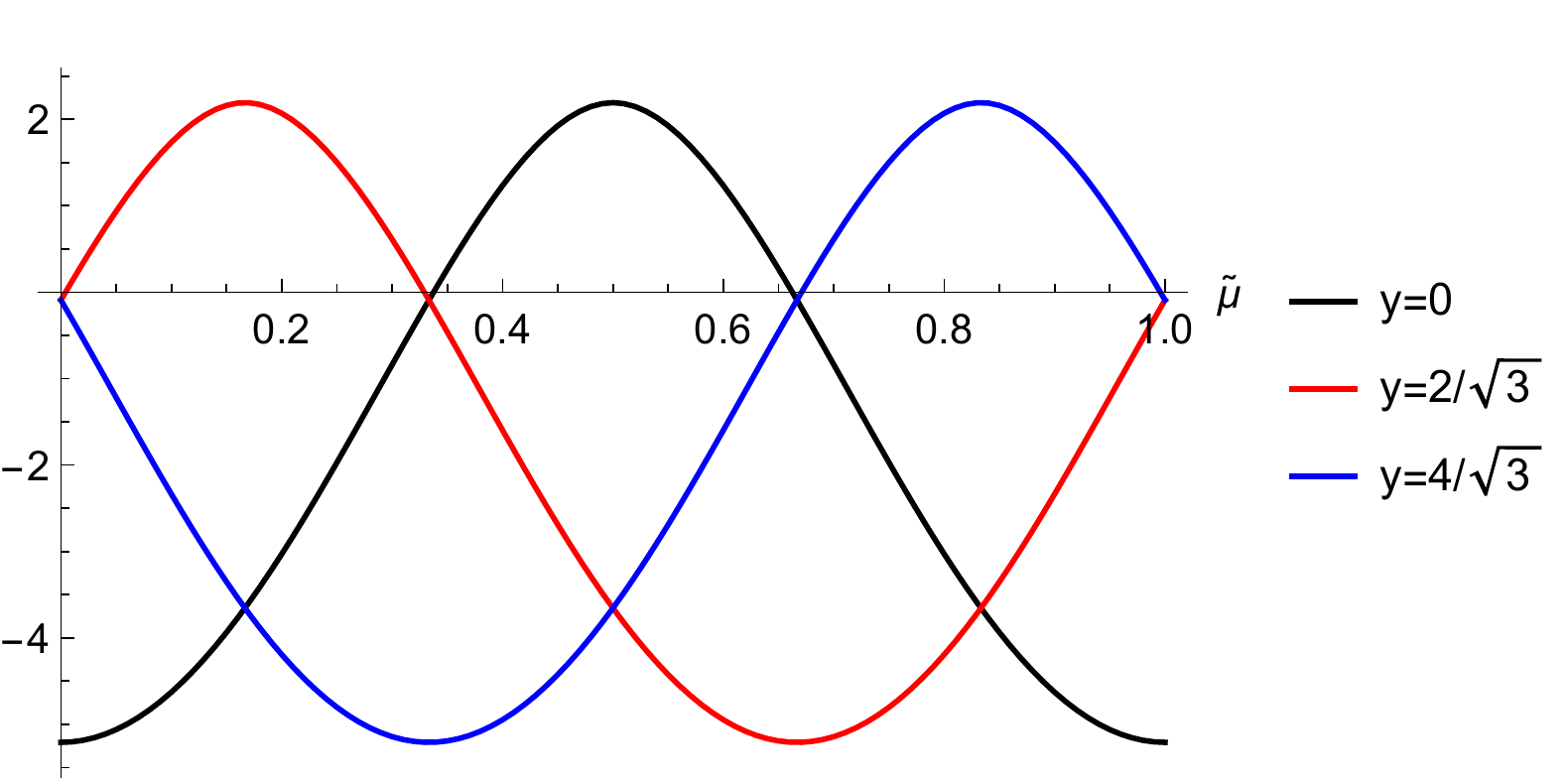}
\includegraphics[width=0.45\textwidth]{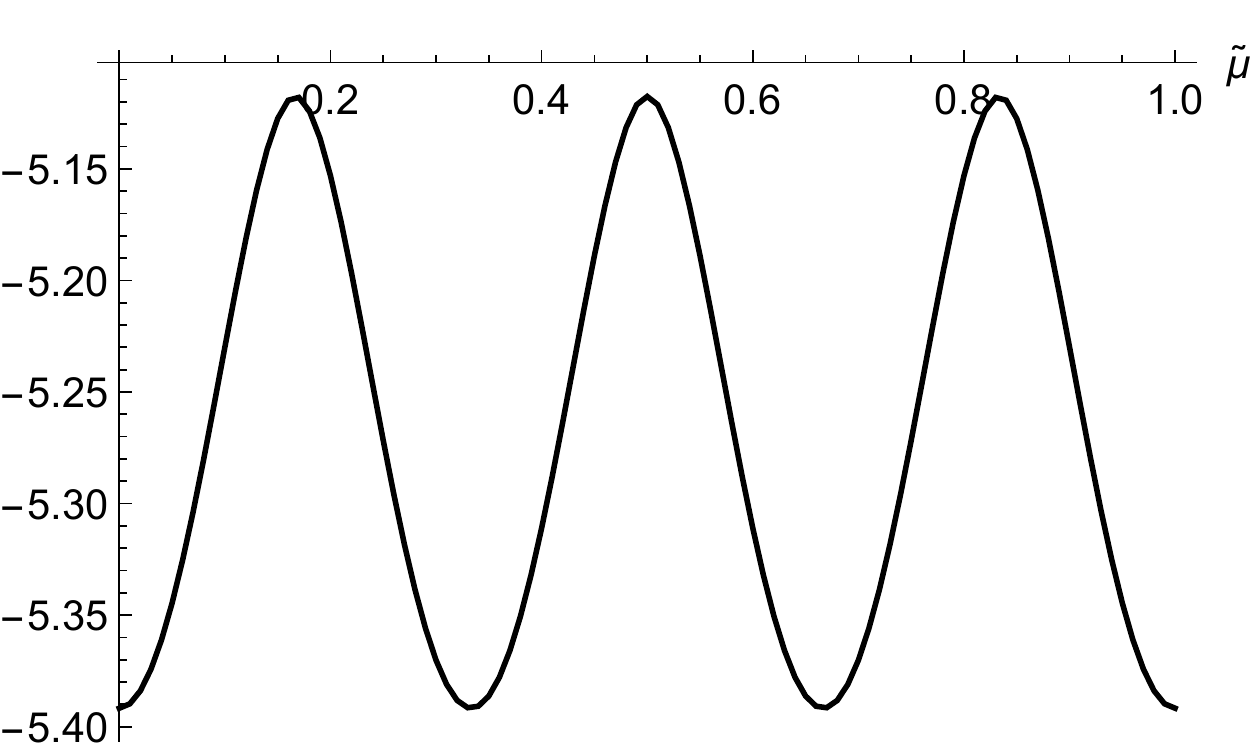}
	\caption{The depths of three consecutive  minima with $g=0$, $N_f=3$ and $x=0$ shown in Fig. \ref{fig:4} as function of imaginary chemical potential $\tilde\mu$ (left panel) and their average (right panel).}
\label{fig:5}\end{figure}

\begin{figure}
	\includegraphics[width=0.45\textwidth]{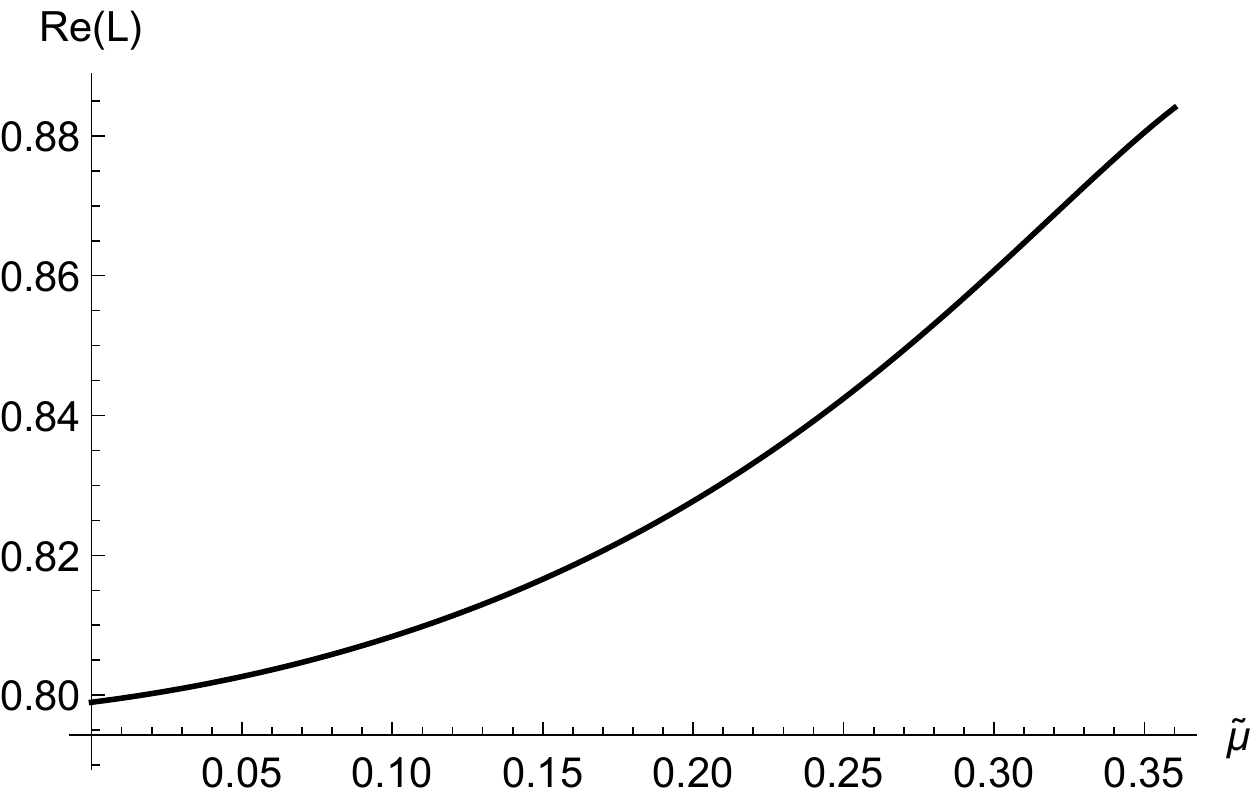}
	\includegraphics[width=0.45\textwidth]{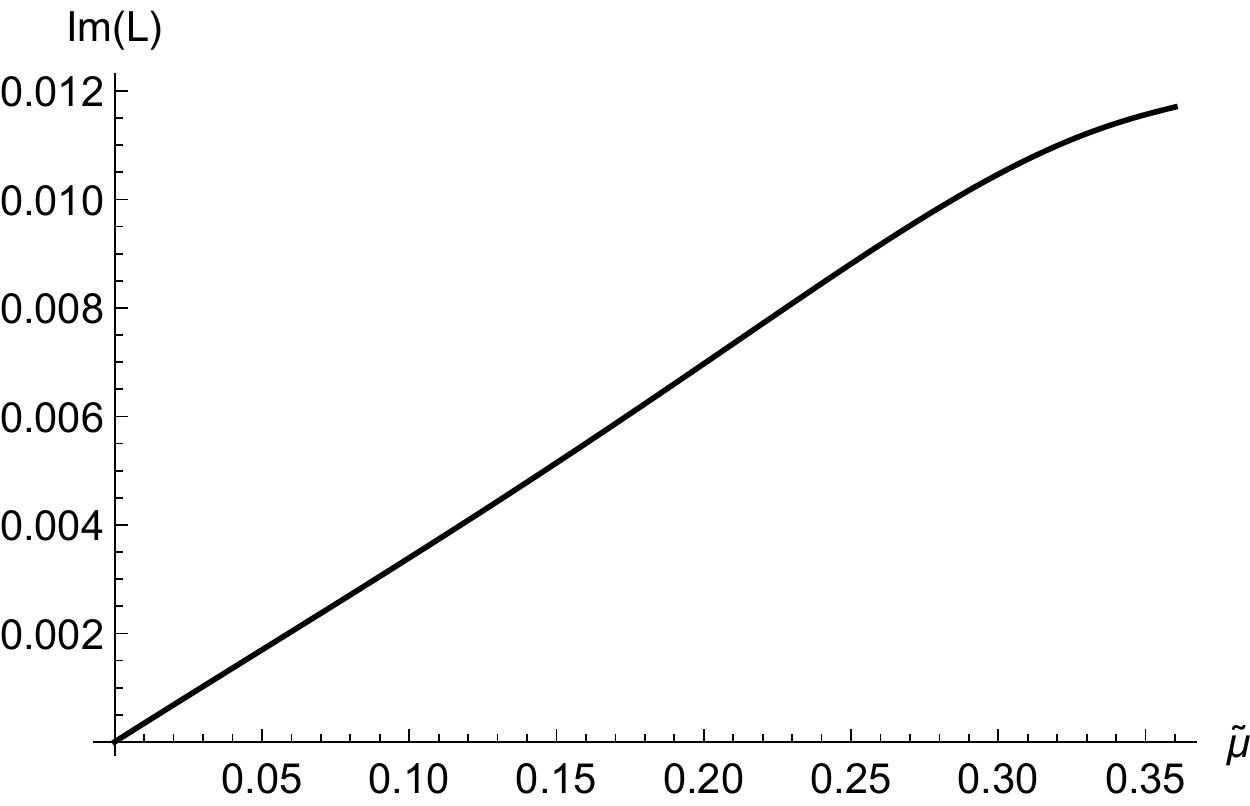}
	\caption{Real (left panel) and imaginary (right panel) parts of the Polyakov loop \Ref{PL} as function of $\tilde\mu$ (imaginary chemical potential). The (small) imaginary part results from the $y$-dependence shown in Fig. \ref{fig:3b}.  The parameters are $g=4$, $N_f=3$.
	}
	\label{fig:L1}\end{figure}
\fi

We turn to the Polyakov loop \Ref{PL}. In the presence of imaginary chemical potential it depends, of course, also on $\mu$. Inserting the numerical $\mu$-dependence, which is shown in Fig. \ref{fig:3b}, into $L$, we can represent this dependence graphically. This is done in Fig. \ref{fig:L1}. We see that the change with respect to eq. \Ref{L2} is weak. An imaginary part shows up, which is small since it results from the $y$-dependence shown in Fig. \ref{fig:3b}. The Fig. \ref{fig:L1} is done for the minimum in the fundamental sector. The picture does not change under the transformations \Ref{slatt}. Under the rotations \Ref{omin}, for odd $k$ the imaginary part changes sign, the real part stays in place.

We complete this section with the following speculation. Assume, the effective potential with imaginary chemical potential had some physical relevance. Then we had, for different $\mu$, equivalent minima. Hence, there should be some tunneling between them. If we take a simple average of the effective potential over these $\mu$, we  get the picture shown in Fig. \ref{fig:5} (right panel). This would be a way to restore the   $Z_3$ symmetry including quarks.
\section{\label{T5}Conclusions and discussion}
In QCD, on  the base of Nielsen's identity and the effective potential of order parameter $W_{L}(A_0^{cl})$ calculated in sects. 2, 3, we  investigated the influence of  chemical potential $\mu$  on the condensation of color constant potentials  $A_0^3$ and  $A_0^8$ at high temperature after the chiral phase transition when quark masses are $m_f = 0$.

For $\mu=0$, in two loop order   we confirm the known results for condensation,  $A_0^3\ne 0$ and   $A_0^8 = 0$ to  order $g^2$, in the main sector, eqs. \Ref{min}, and for the minimum of the effective potential  \Ref{Wmin}, to order $g^4$. We give explicit formulas for all other sectors,  \Ref{omin} and \Ref{slatt}.

For non-zero $\mu$, within the perturbative expansion, we observe no dependence of the condensate on $\mu$. Not expanding in $g$, we observe a weak dependence of the condensates on $\mu$ (see Fig. \ref{fig:3b}). We  see a deepening of the minima on $\mu$ (which is not small); see  Figs. \ref{fig:3a} and \ref{fig:Wmu}..

Within the perturbative expansion, we calculated the Debye mass, \Ref{Dmass}, in the minimum of the effective potential, \Ref{Mdmin}.

 Also we investigated imaginary chemical potential and found  that, including the fermions, there is no $Z_3$-symmetry in our approach{, in accordance with general expectations}.

We obtain that  a  real chemical potential decreases the effective action {and, to some extend, amplifies its dependence of the condensates}. The dependence on imaginary  chemical potential is, of course, periodic. Examples are shown in  Fig. \ref{fig:4} and  Fig. \ref{fig:5}. We mention that the expansion of the  effective potential in powers of $\mu$ has, again in the given approach,  odd contributions.
%
%

{We mention that the effective potential in a minimum is not even in the chemical potential, see for instance the linear term in  \Ref{mmu1}. This could pose an additional problem in attempts of analytic continuation from imaginary to real chemical potential as it was done, for example in \cite{deli03-67-014505}.
 }

We mention that an $A_0$-condensate will be present at least in the perturbative approach, for instance also in a quark-gluon plasma   in the parameter regions where this approach is reliable.  Since this background  violates $Z_3$ and color symmetries, new effective vertexes and related with them  phenomena have to exist. Possibly, these open the possibility for new signals of the deconfinement phase transition.
It is interesting and important that an increase of $\mu$ increases the corresponding effects. Some of these phenomena (such as decay of gluons into photons, etc.) were discussed in \cite{Bord19-134-289}, \cite{Skal19-64-754}.

{In \cite{robe86-275-734}, the $Z_3$-symmetry of the effective potential was discussed. It should have the property $Z(\mu+\frac{2\pi T}{3})=Z(\mu)$, where $\mu$ is the imaginary chemical potential (in \cite{robe86-275-734} the notation $\Theta$ was used),  if only color singlet states contribute to the functional integral \Ref{Z}. In a loop expansion, as we do, this symmetry should be absent. This feature is clearly confirmed by our results, see especially Fig. \ref{fig:4}. The mechanism is that the presence of the quarks makes two third of the gluonic minima (Fig. \ref{fig:2}, right panel) less deep. Now, increasing $\mu$ by $ \frac{2\pi T}{3}$, one cycles through the panels of Fig. \ref{fig:4}. Accordingly, the system moves from one minimum to the next. In Fig. \ref{fig:5}, left panel, the depths of 3 neighbored minima are shown as function of the imaginary chemical potential  $\tilde\mu$. It is seen that at most two of them may be of equal depth. We mention a further symmetry which, possibly, was not observed so far. As function of $x$, for $y=0$, for the effective potential \Ref{Z} the relation
$Z(n+\frac{1}{3}{2\pi T})=Z(n+\frac{2}{3}{2\pi T})$ ($n$-integer) holds.  This property can also be spotted on Fig. \ref{fig:4}.
}
		
It is only   a {\it speculative} tunneling between these minima that could   restore  the $Z_3$ symmetry  as shown in Fig. \ref{fig:5}, right panel. We mention that one observes a different picture  at lower temperature, as is discussed in many lattice investigations, see \cite{deli03-67-014505} for example.

On the base of the obtained gauge-independent effective potential $W_L(A_0^{cl})$, we have calculated thermodynamic pressure $p$ and screening Debye's masses of gluons  $m_D^3$ and $m_D^8$ at the considered environment.

Now, we  compare the results obtained with the calculations existing in the current lite\-rature. In  \cite{guoy19-5-042}, the two-loop constrained  effective potential for QCD at finite temperature was calculated for either massless or massive quarks and chemical potential.  Therein the goal to investigate a spontaneous generation of gauge fields was not intended. Nevertheless some comparisons are of interest and possible. First,  as concerns $\mu$ dependence of the two-loop effective potential calculated in \cite{guoy19-5-042}, it is similar to that derived in \Ref{Wq}. But  no gauge field condensation was detected.
Second, in \cite{guoy19-5-042}, \cite{kort20-101-094025} it was noted that the methods of Nielsen's identity and effective potential of order parameter $W_L(A_0^{cl})$ are an alternative to the constrained effective potential. But in fact they are mutually related. This is because in all of them  the key idea is realized - to calculate an addition part for the two-loop $\xi$-dependent effective potential, which compensates the change of the tree-level Lagrangian followed when the  gauge-fixing term  is introduced.  In Nielsen's identity approach,  this is realized in the form of the special variation of  gluon fields  removing the variation $\xi \to \xi + \delta \xi$ of the gauge-fixing term. There is the set of such orbits (parameterized by an integration constant $\zeta$) in the $(A_0, \xi)$-plane along which the effective potential is independent of $\xi$. In case of $W_L(A_0^{cl})$, the value of $\zeta$ is chosen in such a way  that the $A_0^{cl}$ and $A_0$ belong to Polyakov's loop. This idea was proposed by Belyaev \cite{bely91-254-153}. It is very important that the characteristic and the one-loop  variation of the PL $\Delta \left < L \right > $ are described by the same equation  \cite{skal94-50-1150}. We have investigated this point in detail in Sect. \ref{T3} and explicitly shown that the factor regulating this contribution should be $\sim g^2 (\xi + 1) B_1(a_i)$ instead of  $\sim g^2 (\xi - 3) B_1(a_i)$ as obtained by Belyaev  (and also in  \cite{guoy19-5-042}, \cite{kort20-101-094025} for the constrained effective potential). Actually, just $(\xi - 3)$ factor results, in particular, in the  $A_0^{cl} = 0$ in two-loop order  \cite{bely91-254-153} (same also in  \cite{kort20-101-094025}). This is a special gauge where there is no renormalization of fields in  order $g^2$. Detailed discussion of these (as well as other related with them) peculiarities    were given in our papers \cite{skal94-9-4747},  \cite{skal94-57-324}, \cite{skal94-50-1150} devoted to the Nielsen identity method. In \cite{kort20-101-094025}, the gauge field condensation has been detected in  order $g^3$, as it was assumed in \cite{skal94-50-1150}, \cite{kala93-302-453}.

In \cite{bori20-28-20} and \cite{bori21-965-115332},  the results on condensation of  $A_0^3$ and  $A_0^8$ in plasma with chemical potential have been obtained in the lattice calculations based on a Polyakov loop model. Thus, physical result is the same in all the calculations discussed. It confirms  that the $A_0$ condensation takes place and that it regulates the infrared dynamics of the fields at high temperature. It worth also to mention that the latter results have been obtained as expansion over $g^{-1}$ in a large coupling constant approximation which is reliable at temperatures not much above the deconfinement temperature $T_d$. Therefore these calculations are complementary  to ours. They support our results  when one moves from low  to high temperatures.

We would like to complete with the conclusion that the $A_0$ background has to result in  new interesting phenomena proper to the plasma.
 The presence of chemical potential should amplify these.
{It is to be mentioned that the physical consequences of the $A_0$-condensation, without and with chemical potential, are underrepresented in the modern discussion of quark-gluon plasma. We hope, that the present, detailed investigation will promote such discussions.}

{As an outlook we mention that it would be interesting to see whether the $A_0$-condensation is stable against the inclusion of  magnetic background fields. Possibly, it could even stabilize the magnetic background as discussed recently in \cite{skal2006.05737}.
}

\section{\label{App}Appendix A}
In this appendix we demonstrate that the inclusion of the chemical potential into the calculation of the effective action can be done by analytic continuation in the Bernoulli polynomials. For this, we remind briefly the main steps of the derivation of eq. \Ref{Wq}. The starting point is the fourth momentum component \Ref{mod} in the Feynman rules, which reads
\eq{A.1}{p_4+x+i\mu
}
with $x\to 2\pi T a_i$ for gluons and $x\to2\pi T c_i$ for quarks using the notations \Ref{modq} and \Ref{ci}. The Feynman diagrams involved in the one and two loop orders result in the following sum/integrals (see, for instance, Appendix A in \cite{enqv90-47-291}, or Appendix 2 in \cite{skal94-57-324})
\eq{A.2}{
	H_0 &\equiv
	T\sum_l\int\frac{d^3p}{(2\pi)^3}
	\ln\left((p_4+x+i\mu)^2+p^2\right),
	&H_1&\equiv	
	T\sum_l\int\frac{d^3p}{(2\pi)^3}
	\frac{p_4+x+i\mu}{(p_4+x+i\mu)^2+p^2},
\nn\\	
	H_2&\equiv	
	T\sum_l\int\frac{d^3p}{(2\pi)^3}
	\frac{1}{(p_4+x+i\mu)^2+p^2},
	&H_3&\equiv
	T\sum_l\int\frac{d^3p}{(2\pi)^3}
	\frac{p_4+x+i\mu}{((p_4+x+i\mu)^2+p^2)^2},
}
with the Matsubara frequency $p_4=2\pi T l$ ($l$-integer). The addendum $\frac12$ for the Fermi statistics is included in the $c_i$'s.

In the case $\mu=0$, i.e., without chemical potential, we have the well known expressions
\eq{A.3}{
	H_0&\equiv
	\frac{2\pi^2}{3}T^4B_4\left(\frac{x}{2\pi }\right),
	&H_1&=
	\frac{2\pi}{3}T^3B_3\left(\frac{x}{2\pi }\right),
	&H_2&=
	\frac{1}{2}T^2B_2\left(\frac{x}{2\pi }\right),
	&H_3&=
	-\frac{1}{4\pi}TB_1\left(\frac{x}{2\pi }\right),
}
where $x$ stands for the $c_i$ defined in \Ref{ci}.
We mention that the above expressions suffer from ultraviolet divergences which are to be removed in the usual way.

We continue by carrying out the summation over $l$ in $H_0$, \Ref{A.1}, using the formula
\eq{A.4}{\ln(\eta^2+\Ga^2)+\sum_{l\ne0}\ln\frac{(2\pi l+\eta)^2+\Ga^2}{(2\pi l)^2}
	=\ln\left(2\left(\cosh(\Ga)-\cos(\eta)\right)\right),
}
where we made the sum in the left side converging by adding  a suitable constant. We mention that this formula holds also for complex $\eta$. With $\eta=\frac{x+i\mu}{T}$ and $\Ga=\frac{p}{T}$, and dropping another constant, we rewrite $H_0$ in the form
\eq{A.5}{H_0&=
	\frac{T}{2\pi^2}\int_0^\infty dp\, p^2\,
	\ln\left(2\left(\cosh(\Ga)-\cos(\eta)\right)\right).
}
Starting from this formula, we restricted ourselves to the consideration of $H_0$ since the other $H_i$ can be handled similarly. Next we drop another constant and substitute $p\to p T$ to arrive at
\eq{A.6}{H_0&=
	\frac{T^4}{2\pi^2}\int_0^\infty dp\, p^2\,
	\ln\left(\left(1-z\,e^{-p}\right)
\left(1-z^{-1}\,e^{-p}\right)\right)
}
with $z=e^{(ix-\mu)/T}$. The logarithm in this expression can be written as a sum of two. Because of   convergence, the same holds for the integral. The resulting integrals are, after integration by parts, integral representations of the polylogarithms and we get
\eq{A.7}{H_0&=
	\frac{T^4}{3\pi^2}\left(Li_4(z)+Li_4(z^{-1})\right).
}
It is important to mention that this representation is valid also for  $\mu\ne0$, i.e., for non-zero chemical potential.

Concerning the symmetries we mention that $H_0$, \Ref{A.7}, is periodic under the shift $x\to x+2\pi$ as before. However, the former symmetry under reflection, $x\to -x$, is lost. Now, with $\mu$, $H_0$ is complex and $x\to -x$, which is equivalent to complex conjugation, can be compensated by $\mu\to -\mu$.  As a consequence, the real part of $H_0$ is even in $\mu$ and its imaginary part is odd. In general, these properties can be seen already in  the initial expression \Ref{A.2}, which is, however not finite whereas \Ref{A.7} is finite.

The last step in our discussion is the application of Jonqui\`{e}re's inversion formula,
\eq{A.8}{Li_n\left(e^{(i x-\mu)/T}\right)
	+(-1)^n Li_n\left(e^{(-i x+\mu)/T}\right)
	&=-\frac{(2\pi i)^n}{n!}B_n\left(\frac{x+i\mu}{2\pi T}\right),
}
and we arrive at
\eq{A.9}{H_0 =\frac{2\pi^2}{3}T^4
	B_4\left(\frac{x+i \mu}{2\pi T}\right)
}
in generalization of \Ref{A.3}. Obviously, similar formulas hold for the other $H_i$. From eq. \Ref{A.8} it is seen that the periodic continuation must be done in $x$, i.e., in the real part of the argument of the Bernoulli polynomials. We mention that this is different, although equivalent, to the continuation used in \cite{kort00-61-056007}.
\section{\label{App1}Appendix B}
The Feynman rules for the propagators and for the vertex in the Euclidean space time are,

\paragraph{quark-gluon vertex}
\be \label{b1} \Gamma^d_\mu = - i g \gamma_\mu (t^d)_{a b}, ~~d = (\pi^{\pm}, \pi^0, K^{\pm}, \bar{K}^{\pm}, \eta),\ee

\paragraph{quark propagator}
\be \label{b2} \Delta^{a b}(p) = i~ diag \left( \frac{\hat{p}_1 - m}{p_1^2 + m^2}, \frac{\hat{p}_2 - m}{p_2^2 + m^2}, \frac{\hat{p}_3 - m}{p_3^2 + m^2} \right)_{a b} ,\ee

\paragraph{gluon propagator}
\be \label{b3} D^{\bar{a} a}(p)_{\mu\nu} = \frac{1}{(p^a)^2} \Bigl (\delta^{\bar{a} a}\delta_{\mu\nu} + (\xi - 1 ) \frac{p_\mu^{\bar{a}} p_\nu^{a}}{(p^a)^2}\Bigr), \ee
where $\delta^{\bar{a} a} = 1 $ for such $\bar{a}, a$ as $ (\pi^{\pm}, \pi^0, K^{\pm}, \bar{K}^{\pm}, \eta)$, respectively.
The fourth momentum components are given in  \Ref{mod} and \Ref{modq}.


The periodic Bernoulli polynomials are defined by $B_n(x)\to B_n(x-[\Re x])$, where $[x]$ denotes the integer part and the 'usual'  Bernoulli polynomials are   $B_0(x)=1$, $B_1(x)=\frac12-x$, $B_2(x)=\frac16-x+x^2$, $B_3(x)=\frac{1}{2}x-\frac32 x^2+x^3$, $B_4(x)=-\frac{1}{30}+x^2-2x^4+x^4$.

\end{document}